\documentclass[a4paper,11pt]{article}
\pdfoutput=1 

\usepackage{jheppub} 
\usepackage[utf8]{inputenc}
\usepackage{setspace}

\usepackage{hyperref}
\definecolor{refkey}{gray}{0.45}
\definecolor{labelkey}{RGB}{155,48,48}

\usepackage{mathtools}
\usepackage{physics}
\usepackage{simplewick}
\usepackage[obeyFinal]{todonotes}

\newcommand{\ldS}{l_{dS}}
\newcommand{\nn}{\nonumber}
\newcommand{\be}{\begin{equation}}
	\newcommand{\ee}{\end{equation}}
\newcommand{\ben}{\begin{equation}}
	\newcommand{\een}{\end{equation}}
\newcommand{\bea}{\begin{eqnarray}}
	\newcommand{\eea}{\end{eqnarray}}
\newcommand{\bA}{\begin{array}}
	\newcommand{\eA}{\end{array}}
\newcommand{\bc}{\begin{center}}
	\newcommand{\ec}{\end{center}}

\newcommand{\del}{\partial}

\usepackage{pifont}

\def\mes[#1]{d^{3}{#1}}

\def\del{\partial}

\def\del{\partial}

\reversemarginpar

\definecolor{UI_blue}{RGB}{32, 64, 151}
\definecolor{UI_red}{RGB}{187, 62, 24}
\definecolor{UI_blue2}{RGB}{0, 84, 147}
\definecolor{UI_red2}{RGB}{159, 32, 66}
\definecolor{UI_gray}{RGB}{169, 169, 169}
\definecolor{UI_sepia}{RGB}{112, 66, 20}
\definecolor{UI_bittersweet}{RGB}{254, 111, 94}
\definecolor{UI_emerald}{RGB}{80, 200, 120}
\definecolor{UI_olivegreen}{RGB}{181, 179, 92}
\definecolor{UI_cadetblue}{RGB}{95, 158, 160}
\definecolor{UI_fuchsia}{RGB}{255, 0, 255}
\definecolor{UI_midnightblue}{RGB}{25, 25, 112}
\definecolor{UI_royalblue}{RGB}{0,35, 102}
\definecolor{UI_periwinkle}{RGB}{204, 204, 255}
\definecolor{UI_redorange}{RGB}{255, 83, 73}
\definecolor{UI_brickred}{RGB}{203,65,84}	
\definecolor{UI_forestgreen}{RGB}{34, 139, 34}
\definecolor{UI_tan}{RGB}{210,180,140}	
\definecolor{UI_burlywood}{RGB}{222,184,135}
\definecolor{UI_burlywood}{RGB}{192,64,0}
\definecolor{UI_darkorchid}{RGB}{153,50,204}

\usepackage{titlesec}
\usepackage{verbatim}
\usepackage{fancyhdr}
\usepackage{enumerate}
\hypersetup{colorlinks=true,citecolor=blue,urlcolor=black}
\usepackage{slashed, enumitem}
\usepackage{array}
\newcolumntype{P}[1]{>{\centering\arraybackslash}p{#1}}
\usepackage{anyfontsize}
	\vspace{-11cm}

	\author{Kanhu Kishore Nanda$^{a}$, Somnath Porey$^{a}$, Masataka Watanabe$^{b}$ and Gopal Yadav$^{a,c}$}
\affiliation[a]{\it Chennai Mathematical Institute,\\
	H1 SIPCOT IT Park, Siruseri 603103, India}
	\affiliation[b]{\it Faculty of Science,\\
		The University of Tokyo,
		Tokyo 113-0033, Japan}
\affiliation[c]{\it Instituto de F\'isica, Pontificia Universidad Cat\'olica de Valpara\'iso,\\
	Av. Universidad 330, Curauma, Valpara\'iso, Chile\\}

	\emailAdd{kanhukishore@cmi.ac.in}
	\emailAdd{sporey@cmi.ac.in}
	\emailAdd{max.washton@gmail.com}
	\emailAdd{gopal.yadav@pucv.cl}
	
	\vspace{1cm}

	\abstract{Following \cite{Yadav:2024ray} we embark on a further study of the codimension two holography in de Sitter (dS) space. The $d$ dimensional bulk dS space is sliced by two end of the world $(d-1)$ dimensional branes which are bestowed with either Lorentzian dS or Euclidean AdS metric joining at a common $(d-2)$ dimensional space where the dual CFT lives. Through explicit computations of the holographic entanglement entropy, partition function, and Renyi entropy in both slicings for the bulk dS space emphasizing their similarities as well as differences we provide further evidence for this duality. We also make comments on further generalization of the codimension two holography.}

	\title{Aspects of Codimension two Holography in de Sitter Space}
	
\begin{document}
		\maketitle
		\flushbottom
		\vskip 10pt

\section{Introduction}
Holography has served as a powerful tool to address many questions in the various branches of physics. In string theory, holography was conjectured as a duality between $\mathcal{N}=4$ SYM theory in four dimensions and type IIB string theory on $AdS_5 \times S^5$ \cite{Maldacena:1997re}. For the AdS spacetime, it is well understood where space is emergent dimension (see \cite{Araujo-Regado:2022gvw} where ``time'' is emergent dimension) and for the flat spacetime progress is still happening, see \cite{deBoer:2003vf,Bagchi:2013hja,Laddha:2022nmj,Donnay:2023mrd,Sen:2025oeq,Ghorai:2025ebc,Ghorai:2026qaj,Bagchi:2026bka}. However, holography in de Sitter space is not well understood. There are various proposals such as dS/CFT correspondence \cite{Strominger:2001pn, Maldacena:2002vr, Witten:2001kn} where dual Euclidean CFT lives at the future boundary $\mathcal{I}^+$ of de Sitter space, static patch holography \cite{Susskind:2021omt,Franken:2023pni} with the stretched horizon as a hologram, DS/dS correspondence \cite{Alishahiha:2004md, Alishahiha:2005dj} where the gravity in $d$ dimensional de Sitter space is dual to two IR CFTs which are coupled to each other and dynamical gravity at the central slice in $d-1$ dimensions, higher spin dS/CFT \cite{Anninos:2011ui} etc. These all are codimension one holography in de Sitter space. See also \cite{Dey:2024zjx,Dey:2026lcu,Hikida:2022ltr,Geng:2019ruz,Kawamoto:2023nki} for further development in holography for de Sitter space.

Another version of holography is also the so-called double holography \cite{Takayanagi:2011zk, Fujita:2011fp}, which has been used to discuss the information paradox \cite{Hawking:1975vcx}. In double holography, we put an end-of-the-world brane \cite{Karch:2000ct} and restrict ourselves to certain regions of the geometry. In de Sitter space, double holography and its application to study the information paradox has been done in \cite{Geng:2021dsinfo}. We can further generalize this construction by putting one more EOW brane in bulk\footnote{Wedge holography has application in Multiverse as well, see, e.g., \cite{Yadav:2023qfg},\cite{Aguilar-Gutierrez:2023zoi},\cite{Yadav:2024ray}.}. These two EOW branes form a wedge-like structure, and the gravity living in the wedge is dual to CFT living at the corner of the wedge. For AdS and flat spacetimes as a bulk theory wedge holography was constructed in \cite{Akal:2020wfl, Miao:2020oey} and \cite{Ogawa:2022fhy} respectively.

For de Sitter space as bulk theory, wedge holography was constructed in \cite{Yadav:2024ray}, and there we conjectured that $d$ dimensional de Sitter gravity living in the wedge is dual to $(d-2)$ dimensional CFT living at the corner of the wedge. See \cite{Aguilar-Gutierrez:2023tic, Maulik:2026vmj} for related brane constructions with de Sitter space on the brane whereas AdS gravity in the bulk. Since in our setup, there is dS gravity in the bulk and therefore we name this model as ``DS wedge holography'' or $dSW_d/CFT_{d-2}$ duality. In this paper, we provide the consistency checks of the DS wedge holography proposed in \cite{Yadav:2024ray} by computing pseudo entropy and partition function. We focus on de Sitter space in four dimensions and find that by doing bulk computation of holographic entanglement entropy, one recovers the entanglement entropy of two-dimensional CFT on $S^2$. In particular, when we use dS/CFT to dualize the gravity living on EOW branes, then we find the pseudo entropy which has real and imaginary contributions from the hemisphere and Lorentzian parts consistent with $dS_d/CFT_{d-1}$ results \cite{Narayan:2022afv,Narayan:2023zen,Goswami:2024vfl,Jiang:2023loq,Doi:2023zaf,Doi:2022iyj}. The central charge is imaginary, which suggests that dual theory is nonunitary.  The partition function computation also confirms that central charge of dual CFT is imaginary in $dSW_4/CFT_{2}$.

The construction of \cite{Yadav:2024ray} is based on a de Sitter (dS) slicing of de Sitter space. In this paper, we pursue a new direction as well and propose an alternative construction based on a Euclidean anti de Sitter (EAdS) slicing of de Sitter space. The setup is constructed as follows: the bulk is a de Sitter spacetime bounded by two branes, on each of which the induced metric is that of Euclidean AdS space. Each EAdS brane can then be further dualized by applying the AdS/CFT correspondence, which yields a codimension two holographic description of de Sitter space. We test this proposal by computing the partition function, two point function of scalar field, entanglement entropy, and Rényi entropy. We find the results to be consistent with those in the existing literature. More concretely, our bulk computations of these observables in four dimensional de Sitter space reproduce the corresponding results of a two dimensional Euclidean conformal field theory.

{ In the above cases the EOW branes meet at future boundary of dS space ($I^{+}$) along an $S^{d-2}$ as in AdS wedge holography \cite{Akal:2020wfl, Miao:2020oey}. But when we dualize the gravity living on EOW branes using $DS/dS$ holography then we find that the Euclidean branes are nested latitude spheres $S^{d-1}$ of $S^d$ that never meet. The region between them is a band, which we call a ``{\it $DS/dS$ band}''. Its dual is obtained in two steps. First, gravity in the band localizes to $(d-1)$-dimensional de Sitter gravity with an effective Newton constant. Second, the $DS/dS$ correspondence relates the latter to two CFTs on the central $S^{d-2}$, coupled to $(d-2)$-dimensional de Sitter gravity. Removing the branes reproduces the full $S^d$ and the de Sitter entropy of $dS_d$, i.e.\ nested $DS/dS$.}

The rest of the paper is organized as follows. In the next section \ref{prelim} we review the construction of wedge holography, in particular explaining the setup and showing that the branes can have either Lorentzian dS metric or a Euclidean AdS metric provided the bulk has a suitable metric. In the next two sections we then consider the branes with Lorentzian metric and Euclidean metric in turn. In section \ref{secls} we compute the partition function, holographic entanglement entropy and pseudo-Renyi entropy and show that resulting answers are expected from $d-2$ dimensional CFT viewpoint. In section \ref{seces} we again compute the partition function, pseudo-Renyi entropy along with massless scalar correlation functions and show the similarities and difference with the Lorentzian metric approach of previous section \ref{secls}.  Appendices \ref{appcorr} and \ref{apprenloreuc} extend the analysis of sections \ref{secls} and \ref{seces}. We make some comments on $DS/dS$ Holography in section \ref{secdsds} details of which are relegated to the appendices \ref{apppartfun} and \ref{DS-dS-above-1/lds}. In the final section \ref{secgenoq} we conclude with some open questions.

\section{Preliminaries}
\label{prelim}
In this section we review some basic facts about the codimension two holography in dS space.

Consider the bulk which is $dS_{d}$. We now put two EOW branes which forms a wedge with $(d-2)$ dimensional intersection. Paralleling the AdS/CFT correspondence we posit that there will be a CFT on the $(d-2)$ dimensional intersection space. The zeroth order check for this correspondence to work is to show that the number of generators is the same on either side. A $(d-2)$ dimensional CFT will have $\frac{d(d-1)}{2}$ number of generators. For $dS_{d}$ the number of generators will be $\frac{d(d+1)}{2}$ which is larger than $\frac{d(d-1)}{2}$. However the presence of branes have actually reduced the number of generators. This is easily understood in the embedding space. The $d$ dimensional dS space can be embedded in $d+1$ dimensional Minkowski space. The position of branes then can be recast as being fixed points on some coordinate say $X_d$. The generators $M_{dA}$, which are $d$ in number are now not present. So the total number of generators we are left with in the bulk is given by 
\begin{equation}
  \frac{d(d+1)}{2}- d = \frac{d(d-1)}{2}
\end{equation}  
which now matches the number of generators on the CFT side. Hence the correspondence is between the $dSW_{d}$ and $CFT_{d-2}$ where $W$ stands for the wedge.

In this paper we will focus on the branes having Euclidean AdS gravity or dS gravity. Let us see then what the metric in the bulk is supposed to be in order for the brane to have dS/AdS gravity on it. The bulk action with one brane is given by 
\begin{equation}
	I = \frac{1}{16 \pi G_d} \int  d^dx\sqrt{g} (R-2\Lambda) - \frac{1}{8\pi G_d} \int d^{d-1}x \sqrt{h} (K-T)  \label{dsact2}
\end{equation}
where $2\Lambda =\frac{(d-1)(d-2)}{l^2_{dS}}$ is the cosmological constant, $T$ is the tension, $h$ is the induced metric and $K$ is the extrinsic curvature of the branes. To fix the tension of the branes we use Neumann boundary condition (NBC) on the end of the world (EOW) branes;
\begin{equation}
	K_{ij} - (K-T) h_{ij} =0  \label{bc}
\end{equation}
Taking the trace we obtain
\begin{equation}
	T = \frac{d-2}{d-1} K, \,\,\, K_{ij} = \frac{1}{d-1} K h_{ij} \label{bc2}
\end{equation}
The equation of motion obtained from the action eq.\eqref{dsact2} by varying the metric is given by
\begin{equation}
	R_{M N} - \frac{1}{2} R g_{M N} = -\Lambda g_{M N} \label{eineq}
\end{equation}
where $M,N$ denote the bulk coordinates. We want to find a metric that satisfies the above equation and the boundary condition eq.\eqref{bc} and also gives rise to a $(d-1)$ dimensional AdS/dS metric on the brane. Let us first take the case that the branes admit a dS space. We will take ansatz for the metric to be, with $l_{dS}=1$,
\begin{equation}
	ds^2 = dw^2 + \exp(2f(w))h_{\mu \nu} dx^\mu dx^\nu, \,\,\, \mu,\nu=1,2,\cdots,(d-1) \label{met}
\end{equation}
where $h_{\mu \nu}$ is the metric on the branes satisfying the condition
\begin{equation}
	R_h = (d-1)(d-2)
\end{equation}
where $R_h$ is the Ricci scalar corresponding to $h$. The Ricci scalar for the full metric eq.\eqref{met} is given by
\begin{equation}
	R =\exp(-2f(w)) R_h -2 (d-1) f''(w) -d(d-1)  f'(w)^2=d(d-1)
\end{equation}
A solution to the above differential equation reads
\begin{equation}
	f(w) = \log(\sin(w))
\end{equation}
Thus the bulk metric eq.\eqref{met} can now be written as
\begin{equation}
	ds^2 = dw^2 + l^2_{dS}\sin[2](\frac{w}{l_{dS}})h_{\mu \nu} dx^\mu dx^\nu \label{met2}
\end{equation}
where we have restored the factors of $l_{dS}$. The branes are located at different values of $w$; $w_1,w_2$. We will take $w_2>w_1$. In the embedding coordinates, where the $dS_d$ can be embedded into the $d+1$ dimensional Minkowski space, $M_{d+1}$ such that
\begin{align}
	-X_0^2+ \sum_{i=1}^{d} X_i^2= l^2_{dS} \label{emrel}
\end{align}
eq.\eqref{met2} can be parametrized as
\begin{align}
	X_{d}= l_{dS} \cos(\frac{w}{l_{dS}})
\end{align}
while the rest of the $d$ coordinates satisfy the relation
\begin{align}
	-X_0^2+ \sum_{i=1}^{d-1} X_i^2= l^2_{dS} \sin[2](\frac{w}{l_{dS}}) \label{emrel2}
\end{align}
which describes a $d-1$ dimensional dS space with radius $l_{dS} \sin(\frac{w}{l_{dS}})$. Thus we see that the branes can be embedded into $M_{d+1}$ as constant $X_{d}$ hypersurfaces satisfying the above relation, eq.\eqref{emrel2}.

The bulk metric (\ref{met2}) at the end-of-the-world branes' location satisfies the Neumann boundary condition (NBC), which for two EOW branes labeled by $\beta=1,2$ is given as
\begin{eqnarray}
\label{NBC}
{K}_{ij}^{(\beta)}-({K}^{(\beta)}-T^{(\beta)})\hat{g}_{ij}^{\beta}=0,
\end{eqnarray}
with ${K}_{ij}^{(\beta)}=\frac{1}{2} \partial_w {\hat{g}_{ij}^{\beta}}|_{w= w_{1,2}}$ and ${K}^{(\beta)}=\hat{g}^{ij,\beta}{K}_{ij,(\beta)}$. The extrinsic curvature and its trace for the bulk metric (\ref{met2}) are given as
\begin{eqnarray}
\label{Extrinsic-curvature-dS-bulk}
& & {K}_{ij}^{(\beta)}=\frac{1}{2} \partial_w {\hat{g}_{ij}}|_{w= w_{1,2}}=\frac{1}{l_{dS}} \cot\left(\frac{w_{1,2}}{l_{dS}} \right)\hat{g}_{ij}|_{w= w_{1,2}},\nonumber\\
& &  {K}^{(\beta)}=\hat{g}^{ij,\beta} {K}_{ij}^{(\beta)}=\frac{d-1}{l_{dS}} \cot\left(\frac{w_{1,2}}{l_{dS}} \right).
\end{eqnarray}
Tensions of EOW branes are obtained using the Neumann boundary condition (\ref{NBC}), and  results given in (\ref{Extrinsic-curvature-dS-bulk}) as follows
\begin{eqnarray}
\label{T-dS}
T^{(\beta)}=\frac{d-2}{l_{dS}} \cot\left(\frac{w_{1,2}}{l_{dS}} \right).
\end{eqnarray} 
Note that (\ref{Extrinsic-curvature-dS-bulk}) and (\ref{T-dS}) are written with $\partial_w$ at both branes. This is the outward direction at $Y_2$, while at $Y_1$ the outward normal is $-\partial_w$, so that both $K^{(1)}$ and $T^{(1)}$ flip sign there; the NBC (\ref{NBC}) is insensitive to this overall sign. \\
~\\
Let us derive the action on the brane. We posited that the brane has dS metric on it to obtain the suitable bulk metric. Now we run a consistency check. Given the metric eq.\eqref{met2} we should be able to obtain the Einstein Hilbert action on the brane which must be a dS action with NBC. Substituting the metric eq.\eqref{met2} into eq.\eqref{dsact2} we obtain
\begin{align}
   I &= \frac{1}{16\pi G_d}\int d^dx \sqrt{g} (R-2\Lambda) -\frac{1}{8\pi G_d} \int_{Y_1}  d^{d-1}x \sqrt{h} (K^{(1)}-T^{(1)}) \nonumber \\
   &+\frac{1}{8\pi G_d} \int_{Y_2} d^{d-1}x \sqrt{h} (K^{(2)}-T^{(2)})  \label{dsdsact} \\
   &= \frac{l^{d-3}_{dS}}{16\pi G_d} \int_{w_1}^{w_2} \sin[d-3](\frac{w}{l_{dS}}) dw \int  \sqrt{h} d^{d-1}x \left(R_h -(d-1)(d-2) \right)  \nonumber \\
   &= \frac{l^{d-3}_{dS}}{16\pi G_{d-1}} \int \sqrt{h} d^{d-1}x \left(R_h -(d-2)(d-3) \right) = I_{dS_{d-1}}\,. \label{dsdsact2}
\end{align}
Therefore, we start with $dS_{d}$ gravity in the bulk and obtain $dS_{d-1}$ gravity on the EOW branes, as in brane-world localization of gravity \cite{Karch:2000ct,Karch:2003em}. To derive the above, we have put the branes $Y_i$ at $w_i$ and defined 
\begin{equation}
	\frac{1}{G_{d-1}} = \frac{l_{dS}}{G_d}\int_{w_1}^{w_2} \sin[d-3](w) dw
\end{equation}
and used
\begin{align}
\sin[2](\frac{w}{l_{dS}})(R-2\Lambda) =  \left(R_h- (d-1)\left(d-2-2\sin[2](\frac{w}{l_{dS}})\right)\right), \quad (K-T)= \frac{1}{l_{dS}} \cot(\frac{w}{l_{dS}}) \nonumber
\end{align}
and the formula
\begin{align}
	\int_{w_1}^{w_2} \sin[d-3](w) dw \left(d-2-(d-1)\sin[2](\frac{w}{l_{dS}})\right) = (\sin[d-1](w_2) \cot(w_2)-\sin[d-1](w_1) \cot(w_1) ). \nonumber
\end{align}
Using the above formula then leads to the simplification
\begin{align}
	- (d-1)\left(d-2-2\sin[2](\frac{w}{l_{dS}})\right) +2 \left(d-2-(d-1)\sin[2](\frac{w}{l_{dS}})\right)= -(d-2)(d-3) \nonumber
\end{align}
Till now we considered the brane to have a dS metric. We now want to create slices of dS that looks like Euclidean AdS. Let us see how that will work out. Consider the bulk dS metric of the following form
\begin{align} 
	ds^2= -d\tau^2+ \exp(2f(\tau)) ds^2_{EAdS} \label{met3}\,.
\end{align}
Then we have
\begin{align}
	R= -\exp(-2f(\tau)) (d-1)(d-2)+ 2 (d-1) f''(\tau)+d(d-1)f'(\tau)^2= d(d-1) \label{Req}\,,
\end{align}
whose solution is $f(\tau)= \log(\sinh(\tau))$. So the metric (\ref{met3}) then is
\begin{align}
	ds^2=l^2_{dS}(-d\tau^2+ \sinh[2](\tau)ds^2_{EAdS})  \label{dsadsmet}\,.
\end{align}
Some comments are in order. First, for later purposes the slicing eq.\eqref{met2} will be referred to as the Lorentzian slicing since the branes have a Lorentzian metric on them while the above eq.\eqref{dsadsmet} will be referred to as the Euclidean slicing. Second, the above metric suggests that by taking the branes at constant values of $\tau$ we create an AdS space on the branes. Using AdS-CFT correspondence at the intersection of the branes we have a $d-2$ dimensional CFT. So together we have a dS space in $d$ dimensional bulk dual to a $(d-2)$ dimensional CFT at the intersection of the two branes. Third, in the above metric $\tau=0$ is the tip of the dS, so for the following analysis we will take $\tau>0$. The branes are then situated at $\tau_1,\tau_2$ with $\tau_2 > \tau_1$. Finally in the Euclidean slicing, referring to eq.\eqref{emrel}, we see that the branes are now located in the embedding coordinates at
\begin{align}
	X_{d}= l_{dS} \cosh(\tau)
\end{align}
Substituting the above relation in eq.\eqref{emrel} leads to
\begin{align}
	-X_0^2+ \sum_{i=1}^{d-1} X_i^2= -l^2_{dS} \sinh[2](\tau) \label{emrel3}
\end{align}
which describes a Euclidean AdS space of radius $l_{dS} \sinh(\tau)$.

In this case we will again impose NBC on the branes such that eq.\eqref{bc} and eq.\eqref{bc2} hold. We can derive the action on the brane similar to that in the Lorentzian slicing case. We have
\begin{align}
	I &= \frac{1}{16\pi G_d}\int d^dx \sqrt{g} (R-2\Lambda) +\frac{1}{8\pi G_d} \int_{Y_1}  d^{d-1}x \sqrt{h} (K^{(1)}-T^{(1)}) \nonumber \\
	& -\frac{1}{8\pi G_d} \int_{Y_2} d^{d-1}x \sqrt{h} (K^{(2)}-T^{(2)}) \label{dsadsact} \\
	&=\frac{l^{d-3}_{dS}}{16\pi G_{d-1}}  \int d^{d-1} y \sqrt{h} (R_h+(d-2)(d-3))  \label{dsadsact2} \,,
\end{align}
which clearly is the AdS action with Neumann boundary conditions. Note that the boundary terms in eq.\eqref{dsadsact} has opposite signs compared to eq.\eqref{dsdsact} which can be explained based on the fact that in the case of eq.\eqref{dsdsact} the normal is spatial while for eq.\eqref{dsadsact} the normal points in the time direction.

To obtain finite answers in CFT from the bulk one performs holographic normalization which amounts to adding the following terms to the bulk action \cite{Balasubramanian:1999re} (see \cite{Balasubramanian:2001nb} for the de Sitter analogue)
\begin{equation}
	I_C= \frac{1}{16 \pi G_{d-1}} \int_\Sigma \sqrt{\sigma} \left(2 K_\Sigma + 2 (3-d) -\frac{1}{d-4} R_\Sigma +\cdots \right)
\end{equation}
where $\Sigma$ is the $d-2$ dimensional hypersurface on which CFT lives, $\sigma$ is the metric on $\Sigma$, $K_\Sigma$ is the trace of extrinsic curvature of the hypersurface, and $R_\Sigma$ is the Ricci scalar built out of $\sigma$.

In summary then we have the following statement for the correspondence,
\vspace{0.2cm}
\\
\noindent\fbox{\parbox{\textwidth}{\textbf{The Codimension Two Holography Statement:} 
		\\
	Classical gravity in $d$ dimensional space $\leftrightarrow$ dS/EAdS Gravity on $(d-1)$ dimensional EOW branes $\leftrightarrow$ $(d-2)$ dimensional CFT at the intersection of EOW branes  }}
\vspace{0.2cm}
\\
where the branes can either have dS or AdS metric on them.

Therefore, DS wedge holography states that the holographic dual of $d$ dimensional de Sitter space in the wedge is CFT in $(d-2)$ dimensions. We can write the precise dictionary as follows
\bea \label{eq:dict}
& & {\rm dSW_d/CFT_{d-2}}: \ \ \ \ \ \ \ \ \ \ \ \ Z_{CFT} \sim e^{i I_{dSW_d}}= e^{i I_{dSW_d}^L-I_{dSW_d}^E},
\eea
where $Z_{CFT}$ is the partition function of defect CFT on $S^{d-2}$ and $I_{dSW_d}$ is the gravitational on-shell action computed in the wedge formed by EOW branes. We have ``$i$'' factor in the dictionary of $dSW_d/CFT_{d-2}$ correspondence because the metric on EOW branes is of no boundary type which consists of Lorentzian (with on shell action $I_{dSW_d}^L$) and Euclidean (with on shell action $I_{dSW_d}^E$) parts. 

\section{Lorentzian Slicing}
\label{secls}
In this section we begin our investigation into the codimension two holography. For simplicity we are restricting ourselves to the duality between $dSW_4$ and $CFT_2$. In particular our computations will reveal that the central charge of $CFT_2$ will be imaginary which we will confirm in three different ways- through partition function, calculation of pseudo-entropy and pseudo-Renyi entropy.
\subsection{Partition Function Calculation}
\label{PF-DS4CFT2}
\begin{figure}[h]
  \centering
  \includegraphics[width=.4\linewidth]{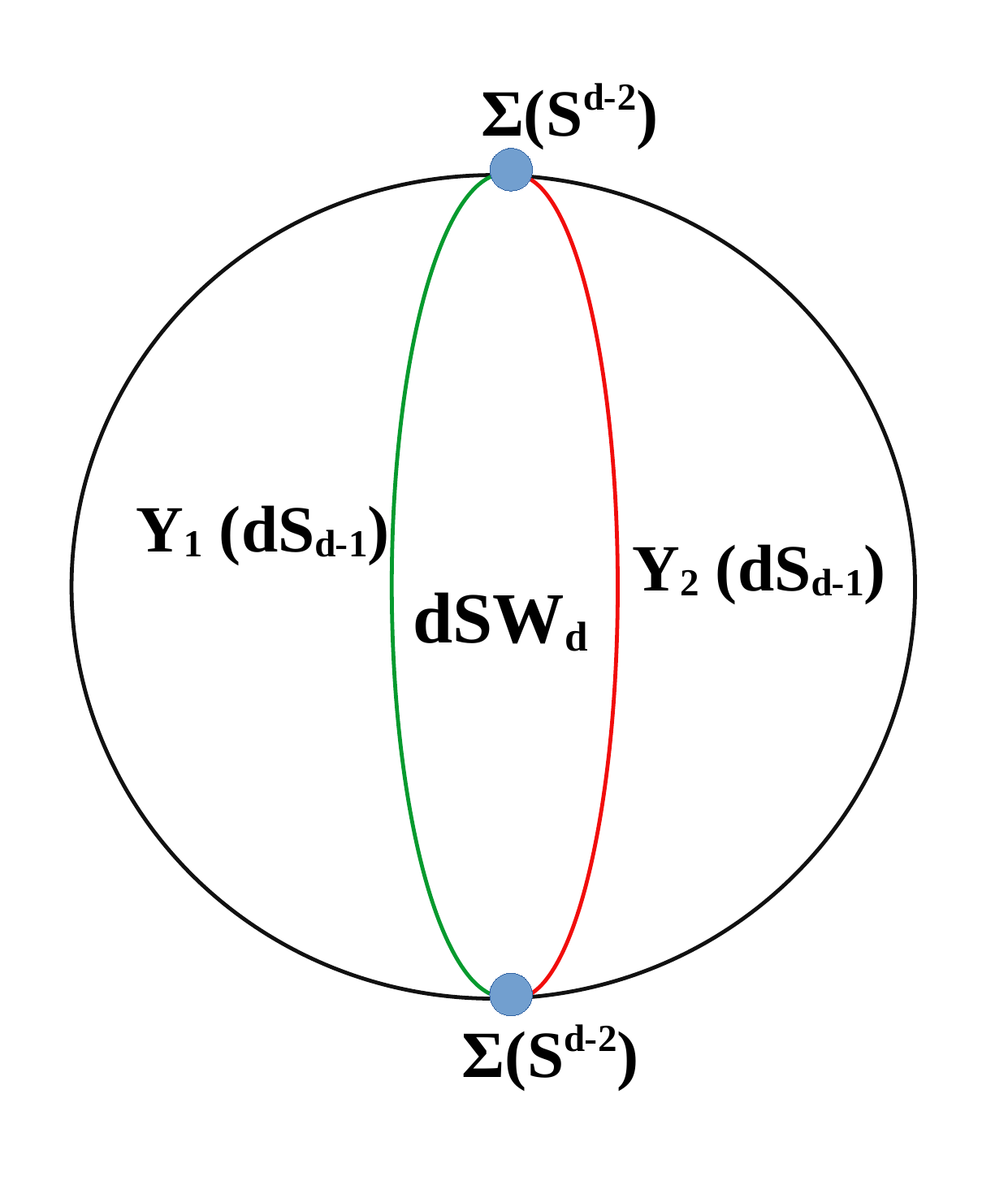}
  \caption{Wedge holography in de Sitter space as a bulk theory: $dSW_d/CFT_{d-2}$ wedge. Green and red curves denote EOW branes at $w=w_1$ ($Y_1$) and $w=w_2$ ($Y_2$). The dual CFT lives on $\Sigma(S^{d-2})$ and is dual to de Sitter gravity in the wedge, ${dSW}_d$.}
  \label{WH-dS-compact}
\end{figure}

In this section, we compute the partition function in DS wedge holography. In particular, we focus on four-dimensional de Sitter space as a bulk theory, and dual CFT is two-dimensional. For this purpose, we need to compute the on-shell action in the wedge geometry, as shown in Fig. \ref{WH-dS-compact}.
Let us rewrite the gravitational action in the bulk, including the boundary terms, as follows
\begin{equation}\label{action-WH}
I_{dSW_d}^L=\frac{1}{16\pi G_d}\int_W\sqrt{-g}\,(R-2\Lambda)-\frac{1}{8\pi G_d}\int_{Y_1}\sqrt{-h}\left(K^{(1)}-T^{(1)}\right)+\frac{1}{8\pi G_d}\int_{Y_2}\sqrt{-h}\left(K^{(2)}-T^{(2)}\right),
\end{equation}
where $Y_{1,2}$ are EOW branes with metric $h$ and $G_d$ is Newton constant in $d$ dimensions, and {$K^{(\beta)}$}, {$T^{(\beta)}$} are given by (\ref{Extrinsic-curvature-dS-bulk}) and (\ref{T-dS}).

Bulk gravity living in the wedge is $DS_d$, and gravity living on EOW branes $Y_{1,2}$ is $dS_{d-1}$. $(d-2)$ dimensional CFT lives at $\Sigma$ located at the cutoff $\tau=\tau_{\infty}$. Future boundary of $dS_{d-1}$[$ds^2=-d\tau^2+l_{dS}^2\cosh^2\left(\frac{\tau}{l_{dS}}\right) d\Omega_{d-2}^2$]) in this case will be $S^{d-2}$. The bulk metric is
\bea \label{metric-dS4-global}
& & ds^2=dw^2+\sin^2\left(\frac{w}{l_{dS}}\right)\left(-d\tau^2+l_{dS}^2\cosh^2\left(\frac{\tau}{l_{dS}}\right) d\Omega_{d-2}^2\right).
\eea
In this case, we have CFT on $S^2$. The gravitational on shell action in four dimensions is given by simplifying eq.\eqref{dsdsact2} for $d=4$, which gives us
\begin{align}
	I^L_{dSW_4} &=\frac{1}{16\pi G_3} \int d\theta d\phi d\tau \sin(\theta) \cosh[2](\frac{\tau}{l_{dS}}) (6-2) = \frac{l_{dS}}{G_3} \int_{0}^{\tau_{\infty}} d\tau \cosh[2](\frac{\tau}{l_{dS}}) \nonumber \\
	&= \frac{l_{dS}}{2G_3} \left(\frac{\tau_\infty}{l_{dS}} + \cosh(\frac{\tau_\infty}{l_{dS}}) \sinh(\frac{\tau_\infty}{l_{dS}})\right)=\frac{l_{dS}}{8G_3} \left(4\frac{\tau_\infty}{l_{dS}} + e^{2\frac{\tau_\infty}{l_{dS}}}\right)   \label{actdswh}
\end{align}
where we have identified
\begin{equation}\label{eq:G3F}
	\frac{1}{G_3}=\frac{\ldS}{G_4}\left(\cos\hat w_1-\cos\hat w_2\right)
\end{equation}
and expanded in large $\tau_{\infty}$. Define the UV cutoff $\epsilon$ in dual CFT on $S^2$ as $\epsilon=2e^{-\frac{\tau_\infty}{l_{dS}}}$ and write on shell action eq.\eqref{actdswh} in terms of $\epsilon$ as below
\begin{align}
   I^L_{dSW_4} = \frac{l_{dS}}{2G_3} \left(\log(\frac{2}{\epsilon}) + \frac{1}{\epsilon^2} \right)  \label{action-WH-v-global-dS4}
\end{align}
The partition function of the dual CFT can be evaluated using $Z \sim e^{i I_{dSW_4}^L}$ as follows
\bea \label{PF-DS4-CFT2}
& &  Z \sim e^{i \frac{l_{dS}}{2 G_3 } \left(\frac{1}{\epsilon^2}-\log(\frac{\epsilon}{2}) \right)}.
\eea
Partition function of two dimensional CFT on $S^2$ with central charge $c$ is \cite{Duff:1993wm} \cite{Henningson:1998gx}
\bea \label{PF-CFT-S2}
& & Z_{CFT} \sim e^{\frac{A}{\epsilon^2}-\frac{c}{3}\log(\frac{\epsilon}{2})},
\eea
where $A$ is some non-universal constant and $\log \epsilon$ is the universal term fixed by the conformal anomaly. Comparing (\ref{PF-DS4-CFT2}) and (\ref{PF-CFT-S2}), central charge is obtained as
\bea \label{central-charge-DS4-CFT2}
& & c= i \frac{3 l_{dS}}{2 G_3 }
\eea  
with, $A=\frac{i\ldS}{2G_3}$. Therefore, the central charge scales with that of $2D$ CFT on $S^2$, and it is imaginary. Thus, we see that dual CFT in $dSW_4/CFT_2$ correspondence is a nonunitary theory.
\subsubsection{Euclidean hemisphere contribution}
According to $dS/CFT$ dictionary, in the bulk, we have no boundary de Sitter space. This involves Lorentzian part glued with Euclidean hemisphere part. We have already calculated above the partition function from the Lorentzian part, here for the completeness, we calculate the hemisphere contribution to the partition function.

The Lorentzian $dS_3$ slice used is given by,
\bea
ds^2_{dS_3} = -d\tau^2 + l_{dS}^2 \cosh^2\ \left(\frac{\tau}{l_{dS}}\right) d\Omega_2^2,
\eea
The above metric is obtained from the round metric on $S^3$ of radius $l_{dS}$,
\bea
ds^2_{S^3} = l_{dS}^2 \, d\theta^2 + l_{dS}^2 \sin^2\theta \, d\Omega_2^2, \qquad \theta \in [0,\pi],
\eea
by the analytic continuation $\theta = \tfrac{\pi}{2} + \tfrac{i\tau}{l_{dS}}$, so that $\sin\theta = \cosh(\tau/l_{dS})$. The complementary Euclidean region, $\theta \in [0,\pi/2]$, is a genuine hemisphere of $S^3$: it is compact, its only boundary is the equatorial $S^2$ at $\theta=\pi/2$ along which it is glued to the Lorentzian part, and no cutoff is needed. The full Euclidean bulk wedge metric is
\bea
ds^2_{E} = dw^2 + \sin^2\!\left(\frac{w}{l_{dS}}\right)\left[l_{dS}^2 \, d\theta^2 + l_{dS}^2 \sin^2\theta \, d\Omega_2^2\right], \qquad \theta \in \left[0,\frac{\pi}{2}\right],
\eea
which is positive-definite.

%

Substituting the above metric in eq.\eqref{dsdsact2}, on shell action for the Euclidean hemisphere is obtained as
\begin{align}
	I^E_{dSW_4} &= -\frac{l_{dS}}{16\pi G_3} 4\int d\Omega_2 \int_{0}^{\frac{\pi}{2}}\sin[2](\theta)d\theta
	= -\frac{\pi l_{dS}}{4G_3}, \label{eq:IE}
\end{align}

The result (\ref{eq:IE}) is half of the Euclidean action $-\frac{\pi\ldS}{2G_3}$ of $S^3$, as expected for the hemisphere. Combining the Lorentzian part of the on shell action with $I^E_{dSW_4}$ above according to (\ref{eq:dict}), we get

\begin{equation}
	iI_{dSW_4}=iI^L_{dSW_4}-I^E_{dSW_4}=\frac{il_{dS}}{2G_3} \left(\log(\frac{2}{\epsilon}) + \frac{1}{\epsilon^2} \right)+\frac{\pi l_{dS}}{4G_3}.
\end{equation}
Hence, partition function in $dSW_4/CFT_2$ turns out to be
\begin{equation}
	Z\sim e^{iI_{dSW_4}}=e^{\frac{il_{dS}}{2G_3} \left(\log(\frac{2}{\epsilon}) + \frac{1}{\epsilon^2} \right)+\frac{\pi l_{dS}}{4G_3}}.
\end{equation}
Thus, we see that the hemisphere gives a finite, real contribution to the partition function, $|Z|\sim e^{\pi\ldS/4G_3}=e^{S_{dS_3}/2}$, as expected for the Hartle-Hawking no-boundary state \cite{Hartle:1983ai}.

\subsection{Pseudo Entropy}
In this section, we compute the pseudo entropy in $dSW_4/CFT_2$ duality. Before proceeding to the actual computation, we first briefly review how one computes the holographic entanglement entropy in AdS wedge holography and then we apply the similar prescription for our purpose. 

In AdS wedge holography  \cite{Akal:2020wfl}, the bulk theory is AdS gravity in $(d+1)$ dimensions, and gravity on EOW branes is AdS type in $d$ dimensions. The latter has been further dualized using $AdS_{d}/CFT_{d-1}$ duality. Overall AdS gravity living in the $(d+1)$ dimensional wedge is dual to $(d-1)$ dimensional CFT, i.e., $Z_{CFT_{d-1}} \sim e^{- I_{AdSW_{d+1}}}$.

To compute the holographic entanglement entropy in AdS wedge holography, authors in  \cite{Akal:2020wfl} consider the $(d-2)$ dimensional subsystem ``A'' [constant $t$ slice] of $(d-1)$ dimensional CFT living at the corner of the wedge ($\Sigma_{d-1}$). The holographic entanglement entropy in AdS wedge holography is calculated using the following formula involving two-step minimization
\bea
& & S_A={\rm Min}_{\gamma_A \ s.t. \ \partial\gamma_{A}^{(1)}= \partial\gamma_{A}^{(2)}= \partial A}\left[{\rm Min}_{\Gamma_A \ s.t. \ \partial \Gamma_A = \gamma_{A}^{(1)} \cup \gamma_{A}^{(2)}}\left(\frac{A(\Gamma_A)}{4 G}\right) \right],
\eea
where $\gamma_{A}^{(1,2)}$ are $(d-2)$ dimensional surfaces on $Y_{1,2}$ such that $\partial\gamma_{A}^{(1)}= \partial\gamma_{A}^{(2)}= \partial A$. To get codimension two surface $\Gamma_A$, i.e., $(d-1)$ dimensional surface in the bulk wedge $AdSW_{d+1}$, we have to extend the $(d-2)$ dimensional compact surface $\gamma_{A}^{(1)} \cup \gamma_{A}^{(2)}$ in such a way that  $\partial \Gamma_A=\gamma_{A}^{(1)} \cup \gamma_{A}^{(2)}$. Minimization in this formula works as follows: fix the shape of $(d-2)$ dimensional surface $\gamma_{A}^{(1,2)}$ and minimize the area of $(d-1)$ dimensional surface $\Gamma_A$ represented by $A(\Gamma_A)$ and then minimize the resulting area by changing the shape of $\gamma_{A}^{(1,2)}$. In this way, one obtains a single minimal surface, and the holographic entanglement entropy [$S_A$ in $(d-1)$ dimensional CFT] is obtained by this minimal area. See Fig. {\bf 5} of \cite{Akal:2020wfl} for pictorial description.

We use a similar procedure for computing the holographic entanglement entropy in DS wedge holography. Here, we use the dS/CFT correspondence from the branes perspective to test the wedge holography in de Sitter space. According to dS/CFT, CFT lives at the future boundary of de Sitter space. Pseudo entropy \cite{Nakata:2020luh} in this context has been computed in the no boundary geometry \cite{Narayan:2022afv,Narayan:2023zen, Goswami:2024vfl} where one obtains contributions from the top-Lorentzian and bottom hemisphere parts. The top-Lorentzian part gives an imaginary contribution, whereas the hemisphere part gives a real contribution. For simplicity we focus on $dSW_4/CFT_{2}$: for four and three-dimensional de Sitter space, pseudo entropy is given by
\be\label{nbdS43rev}
(dS_4)\quad 
S_{nb} = i\,{\pi l_{dS}^2\over 2G_4} {R_c\over l_{dS}} + {\pi l_{dS}^2\over 2G_4}\,;
\qquad\quad
(dS_3)\quad 
S_{nb} = i{l_{dS}\over 2G_3}\log \left({2 R_c\over l_{dS}} \right)
+ {\pi\,l_{dS}\over 4G_3}\,.
\ee
As we are taking $dS_3$ slices of $DS_4$, we expect that pseudo entropy in $dSW_4/CFT_2$ correspondence should come in the above form with the proper identification between Newton constants in three and four dimensions.

\begin{figure}[h]
	\centering
	\includegraphics[width=.6\linewidth]{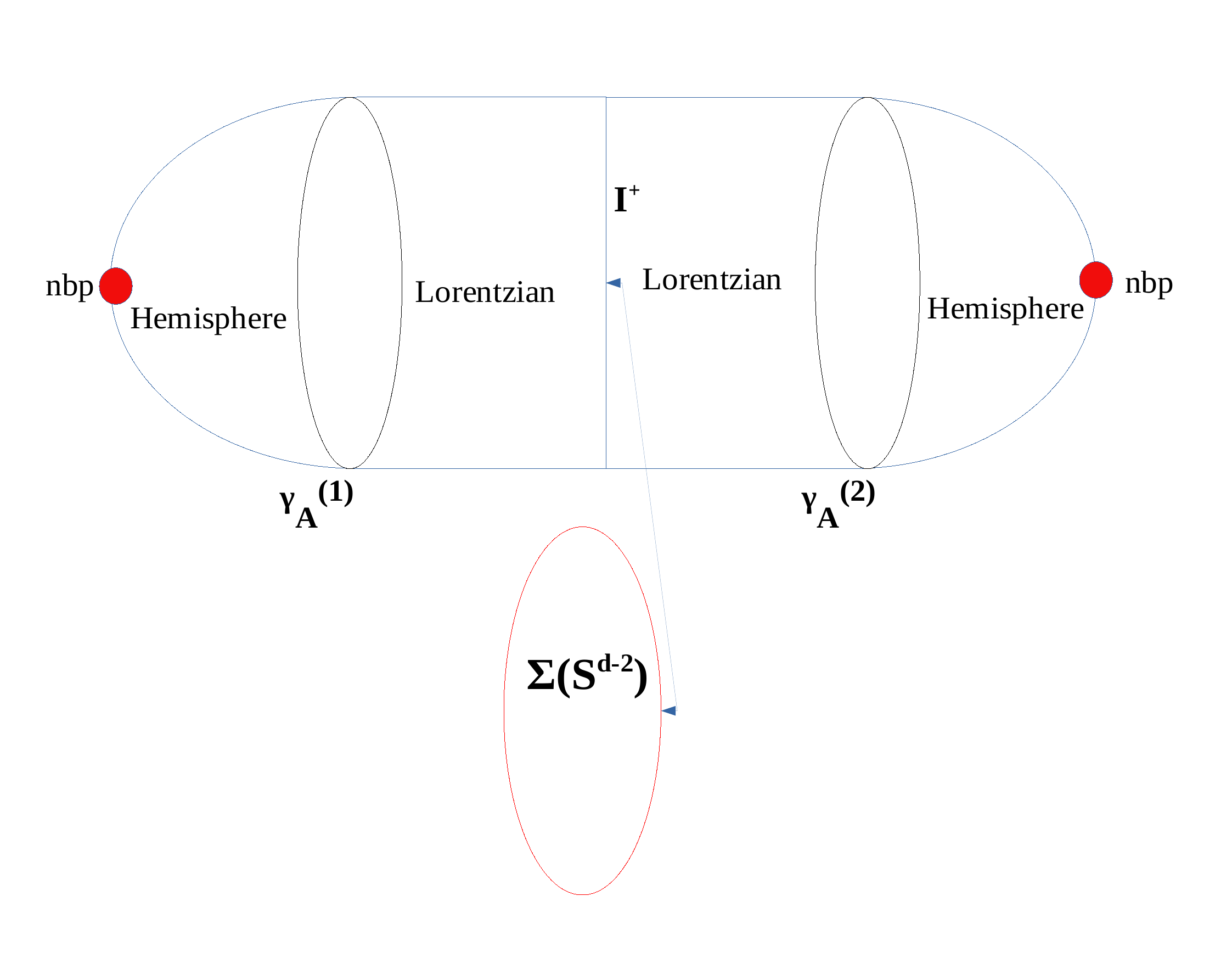}
	\caption{Cartoon of no boundary extremal surfaces in DS wedge holography. This is from the point of view of the wedge formed by $Y_1$ and $Y_2$; we have not shown the bulk extremal surface $\Gamma_A$.}
	\label{WH-dS-compact-ES}
\end{figure}

Let us consider the $d$ dimensional bulk geometry with positive cosmological constant and $dS_{d-1}$ slice as follows
\bea
& & ds_d^2=dw^2+\sin^2 \left(\frac{w}{l_{dS}}\right) ds_{d-1}^2 =dw^2+\sin^2 \left(\frac{w}{l_{dS}}\right)  \left(-d\tau^2+ l_{dS}^2 \cosh^2\left(\frac{\tau}{l_{dS}}\right)\tilde{g}_{ij} dy^i dy^j\right),\nn \\
\eea
where $\tilde{g}_{ij} dy^i dy^j=d\Omega_{d-2}^2$. Consider the extremal surface $M (\Gamma_A)$ as $w(y_i)$ where $y_i$ are the coordinates on $S^{d-2}$. We use the static coordinatization of global de Sitter to compute the pseudo entropy similar to \cite{Narayan:2022afv,Narayan:2023zen,Goswami:2024vfl}. In this representation, the equatorial $\theta_{d-2}=\pi/2$ slice in global de Sitter is equivalent to the constant $t$ slice in static coordinates as described below
\bea \label{induced-metrics}
& & ds^2|_{\theta_{d-2}=\pi/2}^{global}=dw^2+\sin^2 \left(\frac{w}{l_{dS}}\right)  \left(-d\tau^2+ l_{dS}^2 \cosh^2\left(\frac{\tau}{l_{dS}}\right)d\Omega_{d-3}^2\right),\nn \\
& & ds^2|_{t=constt}^{static}=dw^2+\sin^2 \left(\frac{w}{l_{dS}}\right) \left(-\frac{dr^2}{\frac{r^2}{l_{dS}^2}-1} + r^2 d\Omega_{d-3}^2\right),
\eea
using $r = l_{dS}\cosh\left({\tau\over l_{dS}}\right)$. We use the no boundary geometry to compute pseudo entropy in $dSW_4/CFT_2$ case. The no boundary extremal surfaces ($\gamma_A^{(1,2)}$) are shown in Fig. \ref{WH-dS-compact-ES} for $dSW_d/CFT_{d-2}$ correspondence. We anchor the no boundary extremal surfaces at the future boundary ($I^+$) of $dS_{d-1}$ slices in $DS_d$ bulk.
In this case, extremal surfaces $\gamma_{A}^{(1,2)}$ are the no boundary extremal surfaces for the common subregion defined on the future boundary of two branes. The higher dimensional extremal surface $\Gamma_A$, with $\partial\Gamma_A=\gamma_{A}^{(1)} \cup \gamma_{A}^{(2)}$, will start at the no boundary point of one region and end on the subregion $A$; similar for other regions. Overall, $\Gamma_A$ will connect no boundary point of two regions; see Fig. \ref{WH-dS-compact-ES}. Before proceeding further, we would like to write the explicit metric of no boundary geometry in the Lorentzian and hemisphere part in terms of static coordinates, which will be useful later.
\bea \label{induced-metrics-i}
& & {\rm Lorentzian}: \ \ \ \ \ \ ds^2|_{t=constt}^{static}=dw^2+\sin^2 \left(\frac{w}{l_{dS}}\right) \left(-\frac{dr^2}{\frac{r^2}{l_{dS}^2}-1} + r^2 d\Omega_{d-3}^2\right),\nn \\
& & {\rm Hemisphere}: \ \ \ \ \ \ ds^2|_{t=constt}^{static}=dw^2+\sin^2 \left(\frac{w}{l_{dS}}\right) \left(\frac{dr^2}{1-\frac{r^2}{l_{dS}^2}} + r^2 d\Omega_{d-3}^2\right).
\eea
The extremal surfaces from the hemisphere and Lorentzian parts are glued at $r=l_{dS}$.

Now, we discuss the explicit computation for three-dimensional de Sitter gravity on EOW branes, i.e., in $dSW_4/CFT_2$ correspondence. Since the no boundary geometry consists of Lorentzian and hemisphere parts, we divide our discussion into two parts as follows.

{\bf Lorentzian part}: The metric for the constant $t$ slice in the Lorentzian part of four-dimensional de Sitter space is given by
\bea
& & ds^2|_{t=constt}=\Biggl[ l_{dS}^2 d\hat{w}^2+\sin^2\left(\hat{w}\right) \left(-\frac{dr^2}{\frac{r^2}{l_{dS}^2}-1} + r^2 d\theta^2\right)\Biggr],
\eea
where $\hat{w}=\frac{w}{l_{dS}}$. The area of one of the two branches of the codimension two extremal surface, parametrized by $\theta(r)$, is obtained as
\bea
& & A=l_{dS}\int_{\hat{w}_1}^{\hat{w}_2} d\hat{w}\sin\left(\hat{w}\right)\int_{l_{dS}}^{R_c} dr\sqrt{-\frac{1}{\frac{r^2}{l_{dS}^2}-1} + r^2{\theta^{'}(r)}^2}.
\eea
Using the Ryu-Takayanagi formula \cite{Ryu:2006bv, Ryu:2006ef}, the contribution from the Lorentzian part of the geometry to {pseudo entropy} is given by
\bea \label{SEE-DS4}
& & S_4=2\cdot \frac{A}{4 G_4}=\frac{2}{4 G_4}l_{dS}\int_{\hat{w}_1}^{\hat{w}_2} d\hat{w}\sin\left(\hat{w}\right)\int_{l_{dS}}^{R_c} dr\sqrt{-\frac{1}{\frac{r^2}{l_{dS}^2}-1} + r^2{\theta^{'}(r)}^2},
\eea
where the prefactor ``2'' accounts for the two branches of the extremal surface (exactly as in the $dS_3$ computation of \cite{Narayan:2023zen}); 
the $\hat w$ integral already runs over the entire wedge. Now using the relation between Newton constant in three and four dimensions$\left(\frac{1}{G_3}=\frac{l_{dS}}{G_4}\int d\hat{w}\sin\left(\hat{w}\right) \right)$, we can rewrite (\ref{SEE-DS4}) as follows
\bea \label{SEE-DS4-i}
& & S_4=\frac{1}{2 G_3}\int_{l_{dS}}^{R_c} dr\sqrt{-\frac{1}{\frac{r^2}{l_{dS}^2}-1} + r^2{\theta^{'}(r)}^2}\,.
\eea
For the IR subregion, $\theta=\pi/2$, one can show that Lorentzian contribution to the entanglement entropy is given by \cite{Narayan:2023zen}
\begin{equation}
	S_4=\frac{1}{2 G_3}\int_{l_{dS}}^{R_c} dr\sqrt{-\frac{1}{\frac{r^2}{l_{dS}^2}-1} }= i \frac{l_{dS}}{2 G_3}\int_{l_{dS}}^{R_c} dr\sqrt{\frac{1}{r^2-l_{dS}^2} } =i\frac{\ldS}{2G_3}\log\left(\frac{2 R_c}{ l_{dS}}\right). \label{impart}
\end{equation}
In the above computation we chose a particular branch cut that will be shown to be the correct one when we do the Pseudo-Renyi entropy computation.
\\
{\bf Hemisphere part}: For the constant $t$ slice, the metric corresponding to the hemisphere part of the no boundary geometry is given as
\bea
& & ds^2|_{t=constt}=\Biggl[l_{dS}^2 d\hat{w}^2+\sin^2\left(\hat{w}\right)\left(\frac{dr^2}{1-\frac{r^2}{l_{dS}^2}} + r^2 d\theta^2\right)\Biggr].
\eea
The {holographic entanglement entropy} for this case with extremal surface parametrized by $\theta(r)$  is given by
\begin{equation}
	S=2\cdot\frac{A}{4G_4}=\frac{2}{4G_4}\ldS\int_{\hat w_1}^{\hat w_2}d\hat w\sin(\hat w)\int_0^{\ldS}dr\sqrt{\frac{1}{1-\frac{r^2}{\ldS^2}}+r^2\theta'(r)^2},
\end{equation}
where the prefactor ``2'' again counts the two branches of the extremal surface. The entropy contribution from the hemisphere is given as follows
\begin{equation}
	S_4=\frac{1}{2G_3}\int_0^{\ldS}dr\sqrt{\frac{1}{1-\frac{r^2}{\ldS^2}}+r^2\theta'(r)^2}\,.
\end{equation}
For the IR subregion, $\theta=\pi/2$, and hence hemisphere contribution to the {entanglement entropy} is given as follows \cite{Narayan:2023zen}
\begin{equation}
	S_4=\frac{1}{2G_3}\int_0^{\ldS}dr\sqrt{\frac{1}{1-\frac{r^2}{\ldS^2}}}=\frac{l_{dS}}{2G_3}\int_0^{\ldS}dr\sqrt{\frac{1}{l_{dS}^2-r^2}} =\frac{\pi\ldS}{4G_3}.
\end{equation}
Now, the total contributions from the Lorentzian and hemisphere parts, pseudo entropy\footnote{For a detailed calculation of holographic entanglement entropy in $dS_3/CFT_2$, see \cite{Narayan:2023zen}.} in $dSW_4/CFT_2$ holography is given as
\begin{align}
	S_{\rm total}&=i\frac{\ldS}{2G_3}\log\left(\frac{2R_c}{l_{dS}}\right)+\frac{\pi\ldS}{4G_3}=\frac{c}{3}\log\left(\frac{2R_c}{l_{dS}}\right)+\frac{\pi\ldS}{4G_3} \label{eq:Stot}.
\end{align}
Hence, we started with de Sitter gravity in four dimensions and derived the pseudo entropy of two-dimensional Euclidean CFT with $c=i\frac{3 l_{dS}}{2G_3}$. The total hemisphere contribution from both EOW branes is $\frac{\pi l_{dS}}{4G_3}$, which is half of the de Sitter entropy $\frac{2\pi l_{dS}}{4G_3}$ of $dS_3$, as expected for the no-boundary geometry. 

\subsection{Pseudo Renyi Entropy}
\label{lorren}
In this section we follow \cite{Lewkowycz:2013nqa} to obtain the replica geometry suitable to our setup. We will then calculate the  boundary pseudo-Renyi entropy using this replica geometry and show that the pseudo-entropy as written in eq.\eqref{eq:Stot} is correct. Consider then the $S^2$ at the intersection of the two branes with the metric
\begin{align}
	d\Omega_2^2= d\theta^2+\sin[2](\theta) d\phi^2
\end{align}
with $\phi=\phi+2\pi n$. The above spherical metric can be conformally related to a cylinder \cite{Casini:2011kv},
\begin{align}
	d\Omega_2^2= \sin[2](\theta) \left(dx^2+d\phi^2\right), \,\,\, x= \int \frac{d\theta}{\sin(\theta)}
\end{align}
So we take the bulk replica metric to have the following form as it approaches the boundary,
\begin{align}
	ds_b^2 \sim r^2 \left(dx^2+d\phi^2\right)
\end{align}
The replicated bulk metric satisfying the above boundary condition we can then find to be
\begin{align}
	ds^2= dw^2 + l_{dS}^2 \sin[2](\frac{w}{l_{dS}}) \left(-\frac{dr^2}{r^2-r_h^2} + (r^2-r_h^2) dx^2 + r^2 d\phi^2\right) \label{bulkmet}
\end{align}
To fix $r_h$ take the limit $r \rightarrow 0$ in the above metric
\begin{align}
	ds^2= dw^2+  l_{dS}^2 \sin[2](\frac{w}{l_{dS}}) \left(d\rho^2- r_h^2 dx^2+ r_h^2 \rho^2 d\phi^2\right)
\end{align}
So we see that a conical defect develops unless $r_h=\frac{1}{n}$, as for the hyperbolic black holes used in holographic R\'enyi computations \cite{Hung:2011nu}. Now that we have obtained the metric, we can now use it in eq.\eqref{dsdsact2} to find the action which is given by,
\begin{align}
	\tilde{I}_n
	&= \frac{l_{dS}^2}{4\pi G_3} \int dx d\phi \int_{0}^{R_c} r dr=\frac{ l_{dS} }{4G_3} n R_c^2 L_x
\end{align}
where $R_c$ is the cut-off at the future boundary. Using the counter term
\begin{align}
	I_{ct}=\alpha\frac{l_{dS}}{8\pi G_3} \int \sqrt{\gamma} dx d\phi= \frac{\alpha l_{dS} }{4G_3} n L_x R_c\sqrt{R_c^2-r_h^2}= \frac{\alpha l_{dS} }{8G_3} n L_x (2 R_c^2-r_h^2)
\end{align}
we obtain the regularized action to be, with $\alpha=-1$
\begin{align}
	I_n= \frac{ l_{dS} }{8G_3} n r_h^2 L_x= \frac{ l_{dS} L_x}{8G_3 n}
\end{align}
The Renyi entropy, to be compared with the two-dimensional CFT result $S_n=\frac{c}{6}\left(1+\frac1n\right)\log\frac{\ell}{\epsilon}$ \cite{Calabrese:2004eu}, then gives
\begin{align}
	S_n= i\frac{I_n-n I_1}{1-n}=i\frac{l_{dS} L_x}{8 G_3} \frac{1-n^2}{n(1-n)}=  \frac{c}{12} L_x \frac{n+1}{n}
\end{align}
Let us now evaluate $L_x$
\begin{align}
	L_x = \int_{\delta}^{\pi-\delta} \frac{d\theta}{\sin(\theta)}= 2 \log(\frac{2}{\delta})
\end{align}
where $\delta= \frac{l_{dS}}{R_c}$. That takes care of the Lorentzian part. Moving on to the Euclidean contribution with $r \in [0,\frac{1}{n}]$ and $x \rightarrow i t$ we have
\begin{align}
	I^E_n = -\frac{l_{dS}}{4 G_3} \frac{1}{n} \tilde{L}_t
\end{align}
and thus the Renyi entropy is
\begin{align}
	S^E_n = \frac{l_{dS}}{4 G_3}  \tilde{L}_t \frac{n+1}{n}= \frac{n+1}{2n} \frac{\pi l_{dS}}{4 G_3}
\end{align}
where we have taken $\tilde{L}_t= \frac{\pi}{2}$, half of the great circle in the hemisphere. Combining both Lorentzian and Euclidean answers we obtain in the limit $n \rightarrow 1$
\begin{align}
	S= \frac{\pi l_{dS}}{4 G_3}+ \frac{il_{dS}}{2G_3} \log(\frac{2R_c}{l_{dS}}) 
\end{align}
This answer agrees with the standard answer presented in, for example \cite{Nanda:2025tid}.  And the above computation also confirms the choice of branch cut in eq.\eqref{impart}.
\section{Euclidean Slicing}
\label{seces}
In the last section we used the Lorentzian slicing eq.\eqref{met2} as a starting point to explore the duality between $d$ dimensional dS space and $(d-2)$ dimensional CFT. In this section we explore the same duality through a different slicing, namely the Euclidean slicing, eq.\eqref{dsadsmet}, where the branes are bestowed with Euclidean AdS metric. Along with the partition function, pseudo-Renyi entropy computations as in the earlier section, we will also consider scalar fields in the bulk and show that under certain conditions their two point function has the expected CFT form.

\subsection{Correlation Functions of Scalar Fields}
\label{secsc2pt}
In this section we consider a scalar in the bulk dS described by the action
\begin{align}
	I= \int d^d x \sqrt{-g} \left(-\frac{1}{2} g^{MN} \del_M \phi \del_N \phi-\frac{1}{2} m^2 \phi^2\right).
\end{align}
The equations of motion obtained from the above action is
\begin{align}
	(\Box-m^2) \phi=0\,,
\end{align}
which using the metric
\begin{align} 
	ds^2= -d\tau^2+ \sinh[2](\tau) \frac{1}{z^2} (dz^2+dx^a dx_a), \,\,\, a=2,3,\cdots,(d-1) \label{met4}\,,
\end{align}
where we have set $l_{dS}=1$, reads
\begin{align}
	&\frac{1}{\sqrt{g}} \del_M (\sqrt{g} g^{MN} \del_N \phi)- m^2 \phi=0 \nonumber \\
	&\implies -\frac{1}{f(\tau)z^{d-1}}\del_\tau (\sinh[d-1](\tau) f'(\tau)) + \frac{\sinh[d-3](\tau)}{g(z)} \del_z (z^{3-d} g'(z)) -k^2 \frac{\sinh[d-3](\tau)}{z^{d-3}}\\
	& -m^2 \frac{\sinh[d-1](\tau)}{z^{d-1}} =0 \nonumber\,,
\end{align}
where we have taken 
\begin{align}
	\phi(\tau,z,x)= f(\tau) g(z) e^{ik.x} \label{phieq}\,.
\end{align}
Setting $m=0$ we can see that the equation of motion splits into two parts:
\begin{align}
\del_\tau (\sinh[d-1](\tau) f'(\tau))= \varphi^2 f(\tau) \sinh[d-3](\tau) \label{eqf}\,
\end{align}
and
\begin{align}
   \frac{1}{g(z)}  \del_z (z^{3-d} g'(z)) -k^2 \frac{1}{z^{d-3}} - \frac{\varphi^2}{z^{d-1}}=0 \label{geq}\,.
\end{align}
So we see that even though we have taken a massless scalar in the bulk we get a massive scalar on the brane for a generic $\varphi$. In particular we have obtained a spectrum of masses on the brane, $\varphi=\mu l_{dS}$ where $\mu$ is the effective mass and $l_{dS}$ is the dS radius. In what follows we will mostly be interested in setting $\mu=0$ or equivalently $\varphi=0$ which then gives us a massless field on the brane. To find solutions to the above equations we need the following boundary conditions. From the action of the scalar field the boundary terms become
\begin{align}
   I= \frac{1}{2}\int dz d^{d-2} x (\sqrt{-g} \phi \del_\tau \phi)|^{\tau_2}_{\tau_1} - \frac{1}{2}\int d\tau d^{d-2} x \frac{1}{\sinh[2](\tau)}(\sqrt{-g} z^2\phi \del_z \phi)|^{\epsilon}_{\infty} \label{bndact}\,.
\end{align}
The boundary condition at the branes then turns out to be
\begin{align}
	\del_\tau \phi|_{\tau_i}=0 \label{bc3}\,.
\end{align}
We will also take
\begin{align}
	\lim_{z\rightarrow \infty} \phi(k,z) \rightarrow 0 \label{bc4}\,.
\end{align}
Using eq.\eqref{phieq} in eq.\eqref{bc3} leads to $f'(\tau_i)=0$. For $\varphi=0$ solving the equation eq.\eqref{eqf} then gives
\begin{align}
	f'(\tau)= \frac{c_1}{\sinh[d-1](\tau)}\,.
\end{align}
The above equation vanishes at both ends provided $c_1=0$. Thus we get a constant with the aforementioned boundary condition. Then for $\varphi=0$, eq.\eqref{geq} has the solution, with the boundary condition eq.\eqref{bc4}
\begin{align}
	g(z)= z^{\frac{d-2}{2}} K_\nu (k z) \label{gsol}\,,
\end{align}
where
\begin{equation}
	\nu= \frac{d-2}{2}\,.
\end{equation}
Thus for a massless field on the brane we have the following solution:
\begin{align}
	\phi= \kappa z^{\frac{d-2}{2}} K_{\frac{d-2}{2}} (k z) \exp(i k.x)\,.
\end{align}
Up to an overall constant this is the standard solution for a massless scalar field in $d-1$ dimensional AdS space. Thus we can now use the standard AdS tricks to obtain the two point function. The constant $\kappa$ will not matter that much. In fact, following \cite{Freedman:1998tz} the bulk field in terms of the boundary value can be written as
\begin{align}
	\phi(k,z)= \frac{z^{\frac{d-2}{2}} K_{\frac{d-2}{2}} (k z)}{\epsilon^{\frac{d-2}{2}} K_{\frac{d-2}{2}} (k \epsilon)} \phi_B(k)\,.
\end{align} 
We now compute the action eq.\eqref{bndact} using the above scalar field 
\begin{align}
	I&= -\frac{1}{2}\int d\tau d^{d-2} k z^{-d+3} {\sinh[d-3](\tau)} \phi(k,z) \del_z\phi(-k,z)|_{z=\epsilon} \nonumber \\
	&= -\frac{\cosh(\tau_2)-\cosh(\tau_1)}{2}\int \frac{1}{\epsilon}  \frac{d^2 k }{4\pi^2} \frac{1}{\epsilon K_1(k\epsilon)} \phi_B(k) \phi_B(-k)  (\del_z K_1(k z))|_{z=\epsilon}  \nonumber\,,
\end{align}
where we have taken $d=4$ in the final equality. Expanding in $\epsilon$ and keeping the term that is independent of it we obtain,
\begin{align}
 I \sim -\frac{\cosh(\tau_2)-\cosh(\tau_1)}{2} \int \frac{d^2 k }{4\pi^2}k^2 \log(k) \phi_B(k) \phi_B(-k) \,.
\end{align}
Writing $\Psi= \exp(iI)$ and the two point function to be 
\begin{align}
		\expval{O(k) O(-k)}= \frac{\del^2 \log(\Psi)}{\del \phi_B(k) \del \phi_B(-k) }\,, \label{2ptrel}
\end{align}
we finally get
\begin{align}
	\expval{O(k) O(-k)} = -i(\cosh(\tau_2)-\cosh(\tau_1)) k^2 \log(k)\,. \label{sc2pt}
\end{align}
In the position space Fourier transforming the above result and restoring factors of $l_{dS}$ leads to
\begin{align}
	\expval{O(x) O(y)}= -i(\cosh(\tau_2)-\cosh(\tau_1)) \frac{2 l_{dS}}{\pi \abs{x-y}^4}
\end{align}
The above answers agree with eq.(B.23) and eq.(B.24) of \cite{Dey:2024zjx} up to an overall sign. In appendix \ref{appcorr} we present a detailed analysis for the generic $\varphi$ case which reproduces the results of this section for $\varphi=0$.

\subsection{Partition Function}
In this section we want to compute the leading order partition function for $d$ dimensional dS space using the metric 
\begin{align}
	ds^2=l^2_{dS}(-d\tau^2+ \sinh[2](\tau)(dw^2+\sinh[2](w)d\Omega_{d-2}^2))  \label{dsadsmet2}\,,
\end{align}
to obtain the central charge of $(d-2)$ dimensional CFT. To be concrete we will set $d=4$. So we have from eq.\eqref{dsadsact2}
\begin{align}
	I_{dSW_4}^L&= -\frac{l^2_{dS}}{G_4} (\cosh(\tau_2)-\cosh(\tau_1)) \int_{0}^{w_0} dw \sinh[2](w) \nonumber \\ 
	&= \frac{l^2_{dS} }{2G_4} (\cosh(\tau_2)-\cosh(\tau_1)) (w_0-\sinh(w_0)\cosh(w_0))\,.
\end{align}

For $w_0 \gg 1$ we get
\begin{align}
	I_{dSW_4}^L=\frac{l^2_{dS}}{2G_4} (\cosh(\tau_2)-\cosh(\tau_1)) (w_0-\frac{1}{4}\exp(2w_0))\,.
\end{align}
Letting $\frac{2}{\epsilon}=\exp(w_0)$ we obtain
\begin{align}
	I_{dSW_4}^L= \frac{l_{dS}^2}{2G_4}(\cosh(\tau_2)-\cosh(\tau_1)) \left(-\frac{1}{\epsilon^2} + \log(\frac{2}{\epsilon})\right) \,.
\end{align}
Writing
\begin{align}
 \Psi = \exp(i I_{dSW_4}^L)  \sim \exp(-i\frac{A}{\epsilon^2} +\frac{c}{3} \log(\frac{2}{\epsilon}))
\,,
\end{align}
we get the central charge to be
\begin{align}
	c =i\frac{3l^2_{dS}}{2G_4}(\cosh(\tau_2)-\cosh(\tau_1))= i\frac{3l_{dS}}{2G_3}\,,
\end{align}
which agrees with the central charge in eq.\eqref{central-charge-DS4-CFT2} with
\begin{align}
	\frac{1}{G_3}= \frac{l_{dS}}{G_4} (\cosh(\tau_2)-\cosh(\tau_1))\,.
\end{align}
The behaviour of $\Psi \sim \exp(-i A)$ is consistent with the Hartle Hawking wavefunction \cite{Hartle:1983ai} for the expanding branch. This central charge has been calculated in dS/CFT in other methods as well, see for instance \cite{Maldacena:2002vr,Hikida:2022ltr,Dey:2024zjx}.
\subsection{Entanglement Entropy Calculation}
\label{eucren}
In section \ref{lorren} we obtained the Pseudo-Renyi entropy for the case of Lorentzian slicing. In this section we want to repeat the calculation for the Euclidean slicing. Specifically we want to obtain the same central charge following the method of Renyi entropy calculation. In order to do that we first do a conformal transformation on the $S^2$ at the boundary to make it $R \times S^1$ with $\phi$ parametrizing the $S^1$ having the periodicity $2\pi n$, $\phi \equiv \phi+2\pi n$.  The replicated smooth bulk metric is
\begin{align}
	ds^2= l^2_{dS}\left(-d\tau^2+ \sinh[2](\tau)\left(\frac{dr^2}{f(r)} + f(r) dx^2 + r^2 d\phi^2\right)\right)
\end{align}
with
\begin{align}
	f(r)= r^2+r_h^2\,,
\end{align}
and $r_h= \frac{1}{n}$. The derivation of the metric above is same as eq.\eqref{bulkmet}. Substituting the above metric into the action eq.\eqref{dsadsact2} we get
\begin{align}
	\tilde{I}_n 
	&= -\frac{l^2_{dS}}{4\pi G_4} (\cosh(\tau_2)-\cosh(\tau_1)) \int r dr dx d\phi = -\frac{l_{dS}}{8\pi G_3}  L_x\,. R_c^2 2\pi n
\end{align}
To remove the divergent piece we add a counter term
\begin{align}
	I_{ct} = \frac{l_{dS}}{8\pi G_3}L_x 2\pi n (R_c^2 +\frac{r_h^2}{2})\,.
\end{align}
The regularized action adding both contributions is given by
\begin{align}
	I_n= \frac{l_{dS}}{8G_3} n r_h^2 L_x \,.
\end{align}
The Renyi entropy is then found to be
\begin{align}
	S_n &=i \frac{I_n- n I_1}{1-n} = \frac{il_{dS}}{8G_3} \frac{1+n}{n} L_x\,,
\end{align}
where
\begin{align}
	L_x =2\log(\frac{2 R_c}{l_{dS}}) 
\end{align}
In the limit $n \rightarrow 1$ we obtain
\begin{align}
   S_1= \frac{c}{3} \log(\frac{2 R_c}{l_{dS}})  \,.
\end{align}
Thus even in the Euclidean slicing we get the right Lorentzian part as in eq.\eqref{eq:Stot}.

Finally, we can obtain the same answer by following the standard computation involving a minimal surface. This calculation is most simply done in the metric eq.\eqref{met4}. The only difference in this case will be that the action which will be minimized is given by
\begin{align}
	L = i l_{dS} \int \sqrt{dz^2+dx^2} \frac{1}{z}\,.
\end{align}
Then the entanglement entropy is
\begin{align}
	S= \frac{L}{4G_3}=\frac{c}{3} \log(\frac{2l_{dS}}{\epsilon})\,.
\end{align}
We end this section with a few comments. In this section we have seen that three ways of computing the entanglement entropy gives us the same result with the same central charge. Let us note that in the above computation we only considered the Lorentzian dS and as a result we only obtained the imaginary part of the dS entanglement entropy, see \cite{Nanda:2025tid} which we have confirmed for the case of $d=4$. There is an Euclidean part which gives us half of dS entropy which we have not taken into account above since the Euclidean part at any dimension has the universal form of half of total dS entropy in that dimension which we checked in section \ref{secls} and as such the Lorentzian part was the main focus in this section. In appendix \ref{apprenloreuc} we extend this investigation for $dSW_5/CFT_3$ where we again find that the pseudo-entropy calculated through Pseudo-Renyi entropy agrees with the $CFT_3$ answers obtained in \cite{Nanda:2025tid} and moreover the result in Euclidean slicing again agrees with that of the Lorentzian slicing for the imaginary part. Our analysis suggests that the Pseudo-Renyi entropy computation can be done for any dimension and it will agree with the results given in \cite{Nanda:2025tid}. 

\section{$DS/dS$ band}
\label{secdsds}
\begin{figure}
\begin{center}
\includegraphics[width=\linewidth]{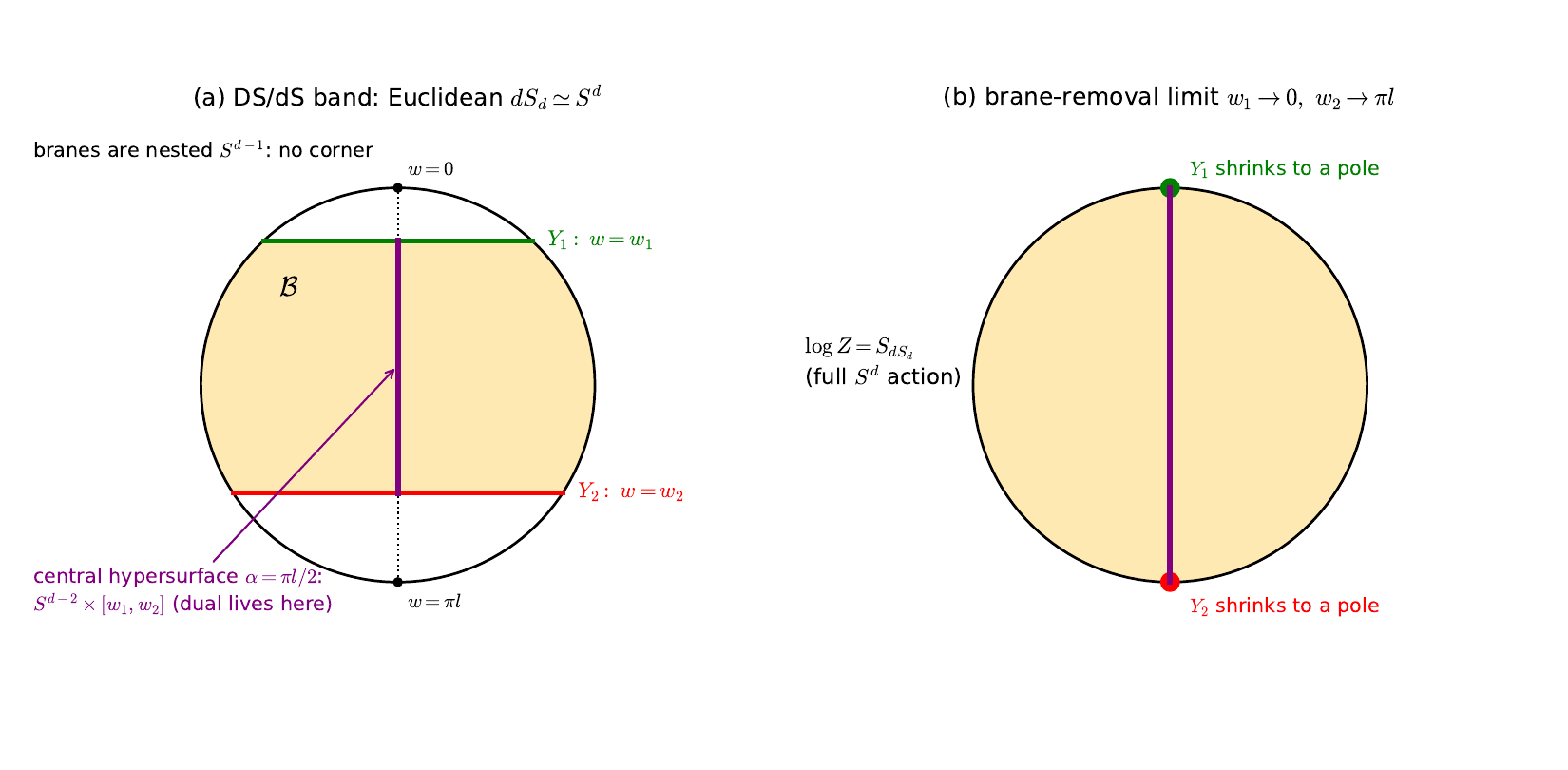}
\caption{The DS/dS band. (a) Cross-section of Euclidean $dS_d\simeq S^d$. The EOW branes $Y_1$ (green) and $Y_2$ (red) are nested latitude spheres $S^{d-1}$. The band $\mathcal B$ (shaded) lies between them and there is no corner. The dotted vertical line is $\alpha=\pi l_{dS}/2$. Its part inside the band (purple), $S^{d-2}\times[w_1,w_2]$ contains the DS/dS central slices of all intermediate spheres where the $(d-2)$-dimensional dual lives after reduction along $w$. (b) Brane removal limit $w_1\to0$, $w_2\to\pi l_{dS}$: the band becomes all of $S^d$ and $\log Z=S_{dS_d}$, eq.~(\ref{band-check}).}
\label{fig:band}
\end{center}
\end{figure}

The bulk metric is
\begin{eqnarray}\label{band-metric}
& & ds^2 =d w^2+ \sin^2\left(\frac{w}{l_{dS}}\right) \left[d \alpha^2+l_{dS}^2\sin^2\left(\frac{\alpha}{l_{dS}}\right) d \Omega_{d-2}^2 \right].
\end{eqnarray}
We denote the $(d-1)$-dimensional EOW branes by $Y_1$ at $w=w_1$ and $Y_2$ at $w=w_2$. Unlike the $dSW_d/CFT_{d-2}$ wedge discussed in Sec. \ref{prelim}, whose Lorentzian branes meet on $\mathcal I^+$, in the Euclidean slicing of (\ref{band-metric}) the constant-$w$ slices are disjoint latitude $(d-1)$-spheres of $S^d$ and do not meet. The region $w_1\le w\le w_2$ is therefore a \emph{band} rather than a wedge, and we call this realization a \emph{DS/dS band} (Fig.~\ref{fig:band}). The bulk is then 
\be
\mathcal B=\{w_1\le w\le w_2\}\simeq S^{d-1}\times[w_1,w_2]\quad(\text{warped}),
\ee
which is a spherical band, not a wedge (Fig.~\ref{fig:band}a). Points of $Y_1$ and $Y_2$ with the same coordinates $(\alpha,\Omega)$ are distinct points of $S^d$, separated by the proper distance $w_2-w_1$.
The holographic duality of $DS/dS$ band is summarized in the following box\\
\vspace{0.2cm}
\noindent\fbox{\parbox{\dimexpr\textwidth-2\fboxsep-2\fboxrule\relax}{%
{\it Classical gravity in the $d$-dimensional DS/dS band $w_1\le w\le w_2$ of Euclidean $dS_d\simeq S^d$}\\[2pt]
\hspace*{2cm}$\updownarrow$ {\footnotesize brane localisation/dimensional reduction \cite{Karch:2003em}, eq.~(\ref{band-Gd})}\\[2pt]
{\it $\equiv$ $(d-1)$-dimensional de Sitter gravity on $S^{d-1}$ with Newton constant $G_{d-1}$}\\[2pt]
\hspace*{2cm}$\updownarrow$ {\footnotesize ``DS/dS correspondence'' \cite{Alishahiha:2004md,Alishahiha:2005dj}}\\[2pt]
{\it $\equiv$ CFT$_L\oplus$CFT$_R$ on the central $S^{d-2}$ ($\alpha=\pi l_{dS}/2$), coupled to $(d-2)$-dimensional de Sitter gravity.}%
}}
We can write precisely the dictionary of $DS/dS$ band as below
\bea \label{eq:dict2}
& &  \ \ \ \ \ \ \ \ \ \ \ \ \ \ \ Z_{CFT} \sim e^{-I_{dS{\mathcal B}_d}},
\eea
where $I_{dS{\mathcal B}_d}$ denotes the Euclidean on-shell action of the band $w_1\le w\le w_2$, and $Z_{CFT}$ is the partition function of the pair of CFTs coupled to $(d-2)$-dimensional de Sitter gravity on the central $S^{d-2}$.
 In the band it sits on the DS/dS central slice of the brane theory: in the bulk, the central slices of all intermediate spheres form the hypersurface $\{\alpha=\pi l_{dS}/2\}\cap\mathcal B\simeq S^{d-2}\times[w_1,w_2]$, and the theory is $(d-2)$-dimensional after the reduction along $w$. Removing the branes ($w_1\to0$, $w_2\to\pi l_{dS}$) turns the band into the full $S^d$, and the second box becomes DS/dS applied twice, $DS_d\to dS_{d-1}\to dS_{d-2}$ (nested DS/dS); see eq.~(\ref{band-check}). In {$DS/dS$} band with four dimensional bulk, the on-shell action is finite, {$\log Z=S_{dS_3}=\pi c/3$}, with real {$c$}.

 In App. \ref{apppartfun}, we computed the partition function of this construction in the energy regime $E \ll \frac{1}{l_{dS}}$ for the four dimensional de Sitter bulk and then generalize to $d$ dimensional bulk. Interestingly by taking the limit such that EOW branes approach the poles of sphere, we recover the partition function of nested $DS/dS$ holography. In App. \ref{DS-dS-above-1/lds}, we computed the partition function for the bulk gravity in the energy regime $\frac{1}{l_{dS}}<E<M_d$ and found the absence of ``log'' term in the on-shell action which suggests that conformal anomaly vanishes in the aforementioned energy regime which is also seen in \cite{Alishahiha:2005dj}. This is, hence, also a complete description of DS wedge holography in the energy regime $\frac {1}{l_{dS}}<E<M_d$. 
%

\section{Discussion}
\label{secgenoq}
We aimed to test the codimension two holography for de Sitter space proposed in \cite{Yadav:2024ray}. As consistency checks of the proposal, we computed pseudo entropy and partition function and showed that these quantities have expected expressions from the dual CFT perspective. We focused on four dimensional bulk and found that holographic computation of pseudo entropy matches the entanglement entropy of CFT in two dimensions. In $dSW_4/CFT_2$ correspondence, the entanglement entropy of two-dimensional CFT (with imaginary central charge) has real and imaginary parts where the real part comes from the hemisphere, and the imaginary part comes from the Lorentzian part of the no boundary geometry. This gives a consistency check of DS wedge holography similar to what has been obtained in AdS wedge holography \cite{Akal:2020wfl,Miao:2020oey}.

We compute the partition function as another quantity to test DS wedge holography. The DS wedge holographic dictionary in $dSW_d/CFT_{d-2}$ correspondence is $Z_{CFT_{d-2}} \sim e^{i I_{dSW_d}}$ as we take no-boundary geometry on EOW branes. The two realizations of DS wedge holography from $dSW/CFT$ and $DS/dS$ perspective on the brane differ geometrically. In $dSW_d/CFT_{d-2}$ the Lorentzian and Euclidean branes (Sec. \ref{prelim}) meet at $\mathcal I^+$ along $S^{d-2}$, as in AdS wedge holography. In $DS/dS$ realization (Sec. \ref{secdsds}), the Euclidean branes are nested latitude spheres of $S^d$ that never meet, and the region between them is a DS/dS band. Its dual follows as: brane localisation to $(d-1)$-dimensional de Sitter gravity with Newton constant (\ref{band-Gd}), followed by DS/dS, which places a pair of CFTs coupled to $dS_{d-2}$ gravity on the central $S^{d-2}$. Removing the branes reproduces the full $S^d$ action and $\log Z=S_{dS_d}$, eq.~(\ref{band-check}), i.e.\ nested DS/dS. In this realization $c$ is identified through $S=\pi c/3$ rather than read off from an anomaly. Only the $dSW_d/CFT_{d-2}$ realization therefore has the geometry of a wedge with an $S^{d-2}$ corner.

Further, we proposed a new way to study wedge holography in de Sitter space in sec. \ref{seces}. Now instead of taking dS slicing of bulk de Sitter space as in \cite{Yadav:2024ray}, we take Euclidean AdS slicing of de Sitter space. By doing holographic computation in the four dimensional de Sitter space, we recover the two point function of two dimensional CFT, correct central charge of CFT in two dimensions from the partition function as well as entanglement entropy.

There are several future directions of DS wedge holography: holographic complexity in DS wedge holography similar to \cite{Yadav:2023sdg} \cite{Aguilar-Gutierrez:2023tic} \cite{Fu:2024vin}; chaos in de Sitter space similar to \cite{Jiang:2024yrk}; fluctuating EOW branes in DS wedge holography similar to \cite{Geng:2022slq}; $T\overline{T}$ deformation \cite{Gorbenko:2018oov}\cite{Coleman:2021nor} in DS wedge holography. It would be interesting to explore these in detail.
\\

\textbf{Acknowledgements:}
We would like to thank Andreas Karch, Ayan Mukhopadhyay, and Tadashi Takayanagi for the very useful discussions; and Andreas Karch, and Tadashi Takayanagi for comments on the older version of the draft. GY is very grateful to Tadashi Takayanagi for hosting him during his visit to the Yukawa Institute for Theoretical Physics. We thank the organizers of the ``Quantum Extreme Universe: Matter, Information, and Gravity'',  ``National Strings Meeting 2024'', and ``STRINGS 2025'' for wonderful hospitality while this work was in progress. KKN, SP and GY would like to thank the Infosys Foundation for the partial support at CMI. MW is supported by a Grant-in-Aid for JSPS Fellows No.~22KJ1777 and by a Grant-in-Aid for Early-Career Scientists No.~25K17387. GY acknowledges support from FONDECYT postdoctoral grant no. 3260698 from Agencia Nacional de Investigación y Desarrollo de Chile. 
\\

\textbf{Statement on the use of AI:} We have benefited from Claude Opus 5.5 version for the geometrical understanding of section 5 in this paper and the figure \ref{fig:band} and also checking the correctness of the calculations in the paper. The physical insights, actual computations, writing of the paper, and the central results are due to the authors and we take full responsibility for any of the errors in the manuscript.
\appendix

\section{Scalar Correlation Function in dS space}
\label{appcorr}
In this appendix we generalize the computation of section~\ref{seces} to a massive bulk scalar and to all the modes along $\tau$, keeping $d=4$. As in that subsection we set $l_{dS}=1$; the factors of $l_{dS}$ can be restored at the end by dimensional analysis, so that $m$ below stands for $m\,l_{dS}$ and the brane mass parameter $\varphi$ for $\mu\, l_{dS}$. The metric in consideration is the Euclidean (EAdS) slicing (\ref{met4}) for $d=4$,
\begin{equation}
	ds^2=-d\tau^2+\sinh^2\tau\,ds^2_{EAdS_3},
	\label{eq:dS4metric}
\end{equation}
where
\begin{equation}
	ds^2_{EAdS_3}=\frac{dz^2+dx^a dx_a}{z^2},\qquad a=2,3 .
	\label{eq:H3metric}
\end{equation}
The two end-of-the-world branes are located at
\begin{equation}
	\tau=\tau_1,\qquad \tau=\tau_2,
	\qquad 0<\tau_1<\tau_2 ,
\end{equation}
and the induced metric on each of them is $EAdS_3$ of radius $\sinh\tau_i$. The conformal boundary of every $EAdS_3$ slice lies at $z\to0$, and this is where the dual CFT$_2$ lives. For later use, we note that the determinant of (\ref{eq:dS4metric}) and the inverse metric components are
\begin{equation}
	\sqrt{-g}=\frac{\sinh^3\tau}{z^{3}},\qquad
	g^{\tau\tau}=-1,\qquad g^{zz}=g^{aa}=\dfrac{z^2}{\sinh^2\tau}.
	\label{eq:detg}
\end{equation}
We consider the same scalar action as in section~\ref{seces}, now for a scalar of mass $m$,
\begin{equation}
	I=-\frac{1}{2}\int d^4x\,\sqrt{-g}
	\left(g^{MN}\partial_M\phi\,\partial_N\phi+m^2\phi^2\right).
	\label{eq:scalarAction}
\end{equation}
Varying \eqref{eq:scalarAction} gives $(\Box-m^2)\phi=0$. Using (\ref{eq:detg}), the Laplacian splits into a $\tau$-part and the Laplacian $\Box_{EAdS_3}$ of the unit $EAdS_3$,
\begin{equation}
	\begin{aligned}\Box\phi&=\frac{1}{\sqrt{-g}}\partial_M\big(\sqrt{-g}\,g^{MN}\partial_N\phi\big)=-\frac{1}{\sinh^3\tau}\partial_\tau\big(\sinh^3\tau\,\partial_\tau\phi\big)+\frac{1}{\sinh^2\tau}\Box_{EAdS_3}\phi,\\ \Box_{EAdS_3}\phi&=z^3\partial_z\big(z^{-1}\partial_z\phi\big)+z^2\partial_a\partial^a\phi,\end{aligned}
	\label{eq:boxsplit}
\end{equation}
so that the equation of motion reads
\begin{equation}
	\left[
	\partial_\tau^2+3\coth\tau\,\partial_\tau
	-\frac{1}{\sinh^2\tau}\Box_{EAdS_3}
	+m^2\right]\phi=0.
	\label{eq:fullKG4}
\end{equation}

As in (\ref{phieq}) we separate variables,
\begin{equation}
	\phi(\tau,z,x)=f(\tau)\,g(z)\,e^{ik\cdot x},
	\label{eq:separation}
\end{equation}
where the $z$- and $x$-dependence solves the Klein--Gordon equation on $EAdS_3$ with a mass parameter $\varphi$ (the separation constant),
\begin{equation}
	\left(\Box_{EAdS_3}-\varphi^2\right)\left(g(z)\,e^{ik\cdot x}\right)=0.
	\label{eq:H3KG}
\end{equation}
Substituting (\ref{eq:separation}) and (\ref{eq:H3KG}) into (\ref{eq:fullKG4}), the $\tau$ equation is
\begin{equation}
	f''+3\coth\tau\,f'
	+\left[m^2-\frac{\varphi^2}{\sinh^2\tau}\right]f=0.
	\label{eq:tauEq}
\end{equation}
Multiplying by $\sinh^3\tau$, it can be written in Sturm--Liouville form,
\begin{equation}
	\partial_\tau\!\left(\sinh^3\tau\,f'\right)
	+m^2\sinh^3\tau\,f
	=\varphi^2\sinh\tau\,f.
	\label{eq:tauSL}
\end{equation}
For $m=0$ this is (\ref{eqf}) with $d=4$, and (\ref{eq:H3KG}) reduces to (\ref{geq}). The quantity $\varphi^2=\mu^2l_{dS}^2$ is therefore the mass of the scalar on the brane, as in section~\ref{seces}.

The boundary condition at the branes follows from the boundary term of the variation of (\ref{eq:scalarAction}). The variation of the scalar action contains
\begin{equation}
	\delta I\big|_{\partial\mathcal M}
	=-\int_{\partial\mathcal M}\sqrt{|\gamma|}\,
	n^M\partial_M\phi\,\delta\phi,
	\label{eq:scalarVariation}
\end{equation}
where $\gamma$ is the induced metric and $n^M$ the unit normal on the boundary $\partial\mathcal M$. At the branes $n^M\partial_M=\pm\partial_\tau$. Requiring (\ref{eq:scalarVariation}) to vanish for arbitrary $\delta\phi$ on the branes, the Neumann condition is
\begin{equation}
	\partial_\tau\phi\big|_{\tau=\tau_i}=0,
	\qquad i=1,2,
	\label{eq:NeumannFull}
\end{equation}
as in (\ref{bc3}), which corresponds to
\begin{equation}
	f'(\tau_1)=f'(\tau_2)=0.
	\label{eq:NeumannF}
\end{equation}
Equations (\ref{eq:tauSL}) and (\ref{eq:NeumannF}) form a Sturm--Liouville problem, so the allowed values of $\varphi^2$ form a discrete set $\varphi_r^2$, $r=0,1,2,\dots$, with eigenfunctions $f_r(\tau)$ orthogonal with weight $\sinh\tau$. To see what fixes the sign of $\varphi_r^2$, we multiply \eqref{eq:tauSL} by $f$ and integrate from $\tau_1$ to $\tau_2$. Integrating the first term by parts produces $\big[\sinh^3\tau\, f f'\big]_{\tau_1}^{\tau_2}$, which vanishes because of \eqref{eq:NeumannF}. We obtain
\begin{equation}
	-\int_{\tau_1}^{\tau_2}d\tau\,\sinh^3\tau\,(f')^2
	+m^2\int_{\tau_1}^{\tau_2}d\tau\,\sinh^3\tau\,f^2
	=\varphi^2\int_{\tau_1}^{\tau_2}d\tau\,\sinh\tau\,f^2.
\end{equation}
Hence,
\begin{equation}
	\varphi^2=
	\frac{
		m^2\displaystyle\int_{\tau_1}^{\tau_2}d\tau\,\sinh^3\tau\,f^2
		-\displaystyle\int_{\tau_1}^{\tau_2}d\tau\,\sinh^3\tau\,(f')^2
	}{
		\displaystyle\int_{\tau_1}^{\tau_2}d\tau\,\sinh\tau\,f^2
	}.
	\label{eq:Rayleigh4D}
\end{equation}
For a massless bulk scalar, $m=0$, every mode has $\varphi^2\le0$, and $\varphi^2=0$ only for the constant mode $f_0=1$. This is the zero mode used in section~\ref{secsc2pt}: it gives a massless scalar on the brane. Non-constant modes with $m=0$ have $\varphi^2<0$.

To solve (\ref{eq:tauEq}) exactly, we define
\begin{equation}
	\rho\equiv\sqrt{\frac94-m^2},
	\qquad
	\nu\equiv\sqrt{1+\varphi^2}, \qquad
	f(\tau)=\frac{Y(\cosh\tau)}{\sinh\tau}.\label{eq:rhonudef}
\end{equation}
With $x\equiv\cosh\tau$ and $\sinh^2\tau=x^2-1$, a short computation turns (\ref{eq:tauEq}) into
\begin{equation}
	(1-x^2)\,Y''-2x\,Y'+\left[\left(\rho-\tfrac12\right)\left(\rho+\tfrac12\right)-\dfrac{\nu^2}{1-x^2}\right]Y=0,
	\label{eq:Legendre}
\end{equation}
where we used $\big(\rho-\frac12\big)\big(\rho+\frac12\big)=2-m^2$. This is an associated Legendre equation of degree $\rho-\frac12$ and order $\nu$, and therefore
\begin{equation}
	f(\tau)=\frac{1}{\sinh\tau}
	\left[
	A_1\,P_{\rho-\frac12}^{\,\nu}(\cosh\tau)
	+A_2\,Q_{\rho-\frac12}^{\,\nu}(\cosh\tau)
	\right].
	\label{eq:Fexact}
\end{equation}
The two Neumann conditions (\ref{eq:NeumannF}) determine $A_1/A_2$ and quantize $\varphi^2$. Note that $\frac32\pm\rho$ are the usual conformal weights of a scalar of mass $m$ in $dS_4$, and that for $m=0$, $\varphi=0$ the constant $f_0=1$ is indeed a solution of (\ref{eq:tauEq}). Note that for $m=0$ and $\varphi^2<-1$ the above solution can be analytically continued to imaginary values of $\nu$. 

We now solve the $EAdS_3$ part (\ref{eq:H3KG}) for a given mode $\varphi=\varphi_r$. With the $x$-dependence $e^{ik\cdot x}$ of (\ref{eq:separation}) and
\begin{equation}
	k^2=k_2^2+k_3^2,
\end{equation}
and using $\Box_{EAdS_3}$ from (\ref{eq:boxsplit}), eq.~\eqref{eq:H3KG} becomes
\begin{equation}
	z^2g''-zg'-(k^2z^2+\varphi^2)g=0,
	\label{eq:radialEq}
\end{equation}
which for $\varphi=0$ coincides with (\ref{geq}) at $d=4$. With the index
\begin{equation}
	\nu=\sqrt{1+\varphi^2},
	\label{eq:nu3}
\end{equation}
already introduced in (\ref{eq:rhonudef}), the substitution $g=z\,\tilde g(kz)$ turns (\ref{eq:radialEq}) into the modified Bessel equation of order $\nu$ for $\tilde g$. The solution regular as $z\to\infty$, i.e. obeying (\ref{bc4}), is
\begin{equation}
	g(z)=\mathcal C(k)\,z\,K_\nu(kz),
	\label{eq:regularSolution}
\end{equation}
where $K_\nu$ is the modified Bessel function of the second kind and $\mathcal C(k)$ is a normalization. For $\varphi=0$, $\nu=1$ and this is (\ref{gsol}) with $d=4$. The $EAdS_3$/CFT$_2$ mass--dimension relation is
\begin{equation}
	\Delta(\Delta-2)=\varphi^2.
	\label{eq:massdimension}
\end{equation}
Hence
\begin{equation}
	\Delta_\pm=1\pm\sqrt{1+\varphi^2}=1\pm\nu.
	\label{eq:Deltapm}
\end{equation}
We take $\Delta=\Delta_+=1+\nu$ as the dimension of the dual operator $\mathcal O$, so that the source has dimension $2-\Delta=1-\nu$. For complex dimensions the correlation functions can be worked out in a coherent state representation, as in \cite{Dey:2024zjx}. In what follows we will focus on real $\nu$ as an illustration.

To read off the source and the response we need the behaviour of (\ref{eq:regularSolution}) near $z=0$. The small-argument expansion of $K_\nu$ for noninteger $\nu$ is
\begin{equation}
	K_\nu(u)=2^{\nu-1}\Gamma(\nu)u^{-\nu}
	+2^{-\nu-1}\Gamma(-\nu)u^\nu+\cdots.
	\label{eq:Kexpansion}
\end{equation}
Using \eqref{eq:regularSolution},
\begin{equation}
	g(z)
	=z^{1-\nu}\left[\phi_B(k)+\cdots\right]
	+z^{1+\nu}\left[\mathcal A(k)+\cdots\right],
	\label{eq:nearBoundary}
\end{equation}
with $\phi_B(k)=2^{\nu-1}\Gamma(\nu)k^{-\nu}\,\mathcal C(k)$ and $\mathcal A(k)=2^{-\nu-1}\Gamma(-\nu)k^{\nu}\,\mathcal C(k)$. The source $\phi_B(k)$ is the coefficient of $z^{2-\Delta}=z^{1-\nu}$, whereas the normalizable coefficient $\mathcal A(k)$ multiplies $z^\Delta=z^{1+\nu}$. Therefore, normalizing to $\phi_B(k)$ gives
\begin{equation}
	\boxed{
		\frac{\mathcal A(k)}{\phi_B(k)}
		=2^{-2\nu}\frac{\Gamma(-\nu)}{\Gamma(\nu)}k^{2\nu}.
	}
	\label{eq:AoverJ}
\end{equation}
Equivalently, in terms of $\mathcal C(k)$,
\begin{equation}
	g(z) \approx \mathcal C(k) \left[ \frac{\Gamma(\nu)}{2} \left(\frac{k}{2}\right)^{-\nu} z^{1-\nu} + \frac{\Gamma(-\nu)}{2} \left(\frac{k}{2}\right)^\nu z^{1+\nu} \right],
	\label{eq:psiSmallZ}
\end{equation}
and, since we identify the source through $g(\epsilon)\simeq \epsilon^{1-\nu}\phi_B(k)$ at the cutoff $z=\epsilon$, as in section~\ref{secsc2pt},
\begin{equation}
	g(z) = \phi_B(k) \left[ z^{1-\nu} + \frac{\Gamma(-\nu)}{\Gamma(\nu)} \left(\frac{k}{2}\right)^{2\nu} z^{1+\nu} +\cdots\right].
	\label{eq:psiSourceNormalized}
\end{equation}
Similarly, for integer $\nu$ ($\nu = n \in \mathbb{Z}^+$) the expansion of $K_n$ contains a logarithm,
\begin{equation}
	K_n(u)=\frac12\sum_{j=0}^{n-1}(-1)^j\frac{(n-j-1)!}{j!}\Big(\frac u2\Big)^{2j-n}+(-1)^{n+1}\frac{1}{n!}\Big(\frac u2\Big)^{n}\log\Big(\frac u2\Big)+\cdots,
	\label{eq:Knexpansion}
\end{equation}
and we can write
\begin{equation}
	g(z) = \phi_B(k) \left[ z^{1-n} +\cdots+ (-1)^{n+1} \frac{1}{2^{2n-1} \, n!(n-1)!} \, k^{2n} \, z^{1+n} \log(kz) + \mathcal{O}(z^{1+n}) \right],
	\label{eq:psiIntegerExpansion}
\end{equation}
where the dots denote the terms $z^{3-n},\dots,z^{n-1}$, which are analytic in $k^2$ and only give contact terms.

The on-shell action reduces to the boundary terms (\ref{bndact}). The brane terms vanish by (\ref{eq:NeumannFull}), and the term at $z\to\infty$ vanishes by (\ref{bc4}). Using (\ref{eq:detg}), $\sqrt{-g}\,g^{zz}=\sinh\tau/z$, so for a single mode $\phi=f_r(\tau)\,g(z)\,e^{ik\cdot x}$ the on-shell action at $z=\epsilon$ is
\begin{equation}
	I_{\rm os}
	=-\frac{1}{2}
	\left(\int_{\tau_1}^{\tau_2}d\tau\,\sinh\tau\,f_r^2\right)
	\int\frac{d^2k}{(2\pi)^2}\,
	\frac{1}{\epsilon}\,
	g(\epsilon,k)\,\partial_zg(\epsilon,-k).
	\label{eq:SosScalar}
\end{equation}
The $\tau$-integral is the norm of the mode with the Sturm--Liouville weight $\sinh\tau$ of (\ref{eq:tauSL}). Defining $N_r=\int_{\tau_1}^{\tau_2}d\tau\,\sinh\tau\,f_r^2(\tau)$,
\begin{equation}
	I_{\rm os}^{(r)}
	=-\frac{N_r}{2}
	\int\frac{d^2k}{(2\pi)^2}\,
	\frac{1}{\epsilon}\,
	g(\epsilon,k)\,\partial_zg(\epsilon,-k).
	\label{eq:SosNorm}
\end{equation}
Inserting (\ref{eq:psiSourceNormalized}), the non-local part of $\frac1\epsilon\, g\,\partial_z g$ comes from the cross terms between $z^{1-\nu}$ and $z^{1+\nu}$ and is proportional to $\mathcal A(k)/\phi_B(k)$ of (\ref{eq:AoverJ}). Treating the cutoff as in \cite{Freedman:1998tz}, which multiplies this term by $2\nu$, and using $\Psi=\exp(iI)$ and $\langle\mathcal O(k)\mathcal O(-k)\rangle=\frac{\delta^2\log\Psi}{\delta\phi_B(k)\,\delta\phi_B(-k)}$ as in eq.\eqref{2ptrel}, we find \textbf{for noninteger $\nu$:}
\begin{equation}
	\langle\mathcal{O}(k)\mathcal{O}(-k)\rangle
	\propto
	-i\,N_r\,
	(2\nu)2^{-2\nu}
	\frac{\Gamma(-\nu)}{\Gamma(\nu)}
	k^{2\nu},
	\label{eq:momentum2pt}
\end{equation}
\textbf{for integer $\nu=n$}, using (\ref{eq:psiIntegerExpansion}) instead:
\begin{equation}
	\langle \mathcal{O}(k) \mathcal{O}(-k) \rangle \propto -i N_r \left[ \frac{(-1)^{n+1} 2^{2-2n}}{((n-1)!)^2} \right] k^{2n} \log k .
	\label{eq:integer2ptFinal}
\end{equation}
The integer result is also the $\nu\to n$ limit of (\ref{eq:momentum2pt}): near $\nu=n$, $\Gamma(-\nu)$ has a pole with residue $(-1)^{n+1}/n!$, and $k^{2\nu}=k^{2n}\left(1+2(\nu-n)\log k+\dots\right)$, which reproduces the coefficient in (\ref{eq:integer2ptFinal}). In the massless case the relevant mode is the zero mode $f_0=1$ with $\varphi=0$, $\nu=n=1$, therefore
\begin{equation}
	N_0=\int_{\tau_1}^{\tau_2}d\tau\, \sinh{\tau}=\cosh{\tau_2}-\cosh{\tau_1},
\end{equation}
and (\ref{eq:integer2ptFinal}) gives $\langle\mathcal O(k)\mathcal O(-k)\rangle=-i(\cosh\tau_2-\cosh\tau_1)\,k^2\log k$, in agreement with eq.\eqref{sc2pt}.
\section{Renyi Entropy in $dSW_5/CFT_3$}
\label{apprenloreuc}
In this appendix we extend the analysis of sections \ref{lorren} and \ref{eucren} by considering the boundary Renyi entropy in the case of $dSW_5/CFT_3$. We will be brief in our presentation since the analysis will be similar to that in sections \ref{lorren} and \ref{eucren} and \cite{Hung:2011nu,Nanda:2025tid}. Note that for both the Lorentzian slicing eq.\eqref{metric-dS4-global}, and Euclidean slicing eq.\eqref{dsadsmet2} the boundary is an $S^3$ which can be shown to be conformally equivalent to $S^1 \times H_2$ as follows:
\begin{align}
	d\Omega_3^2&= d\theta^2 + \sin[2](\theta) d\phi^2+ \cos[2](\theta)d\psi^2= \sin[2](\theta) \left(d\phi^2+ \frac{d\theta^2+ \cos[2](\theta)d\psi^2}{\sin[2](\theta)}\right) \nonumber \\
	&=\sin[2](\theta)(d\phi^2+ dH_2^2)
\end{align} 
with $\phi=\phi+2\pi n$. So we want to find a smooth replicated bulk metric which has the boundary value
\begin{align}
	ds_b^2 \sim r^2 (d\phi^2+ dH_2^2)
\end{align}
Let us first take the Lorentzian slicing. The bulk metric that has the above boundary value, in the Lorentzian slicing case, is given by
\begin{align}
	ds^2= dw^2+ l^2_{dS} \sin[2](\frac{w}{l_{dS}}) \left(-\frac{dr^2}{f(r)} + f(r) d\phi^2+ r^2 dH_2^2\right)
\end{align}
with
\begin{align}
	f(r)= r^2+ 1 + \frac{c_2}{r}
\end{align}
As in \cite{Nanda:2025tid} the horizon $r_h$ is located in the complex plane, $r_h=i \rho_h$ and ensuring that there is no conical deficit leads to
\begin{align}
	c_2= -r_h(r_h^2+1), r_h= \frac{i}{3n} (\sqrt{1+3n^2}+1)
\end{align}
Substituting the above metric in the action eq.\eqref{dsdsact} gives
\begin{align}
	\tilde{I}_n= \frac{l^2_{dS}}{16\pi G_4} \int r^2 dr d\phi dH_2 (R_h- 6)
	=\frac{l_{dS}^2}{4G_4} A n (R_c^3 - r_h^3) \nonumber 
\end{align}
The counter terms needed to the action finite in this case reads
\begin{align}
	I_{ct} &=\frac{l^2_{dS}}{8\pi G_4} \int dH_2 d\phi \sqrt{\gamma} (\alpha R_b + \beta)= \frac{n l^2_{dS}}{4G_4} A R_c^2 \sqrt{R_c^2 +1+ \frac{c_2}{R_c}} (\alpha R_b + \beta) \nonumber \\
	&=\frac{n l^2_{dS}}{4G_4} A \left(-2\alpha R_c + \beta R_c^3 + \frac{\beta}{2} R_c + \frac{\beta c_2}{2}\right)
\end{align}
where we have used $R_b=-\frac{2}{R_c^2}$ and expanded the square root in large $R_c$ limit keeping only the leading terms. Adding the two contributions and canceling $R_c$ dependent terms give
\begin{align}
	\beta= 4 \alpha,  \beta=-1
\end{align}
and,
\begin{align}
	I_n= -\frac{n l^2_{dS}}{8G_4} A (2r_h^3+ c_2)= -\frac{n l^2_{dS}}{8G_4} A r_h(r_h^2-1)
\end{align} 
Computing the Renyi entropy we obtain in the limit $n \rightarrow 1$
\begin{align}
	S_1= \lim_{n \rightarrow 1} i \frac{I_n-n I_1}{1-n}= -\frac{l^2_{dS}}{4G_4} A \label{impart2}
\end{align}
where, as in \cite{Nanda:2025tid}
\begin{align}
	A= -2\pi \left(1+\frac{i R_c}{l_{dS}}\right) 
\end{align}

Let us now move on to the Euclidean slicing where the replicated smooth bulk metric is given by
\begin{align}
	ds^2=l_{dS}^2\left(-d\tau^2+\sinh[2](\tau) \left(\frac{dr^2}{f(r)} + f(r) d\phi^2+r^2 dH_2^2\right)\right)
\end{align}
with
\begin{align}
	f(r)=r^2-1+\frac{c_3}{r}, \,\,\, c_3= r_h(1-r_h^2), \,\,\, r_h= \frac{1+\sqrt{1+3n^2}}{3n} 
\end{align}
Substituting the above metric into the action eq.\eqref{dsadsact2}
\begin{align}
	\tilde{I}_n= \frac{l^{2}_{dS}}{16\pi G_{4}}   \int r^2 dr d\phi dH_2 (R_h+6) = -\frac{n l^{2}_{dS}}{4 G_{4}}  \tilde{A} (R_c^3-r_h^3)
\end{align}
Adding the counter terms
\begin{align}
	I_{ct} &=\frac{l^2_{dS}}{8\pi G_4} \int dH_2 d\phi \sqrt{\gamma} (\alpha R_b + \beta)=\frac{n l^2_{dS}}{4G_4} \tilde{A} \left(-2\alpha R_c + \beta R_c^3 - \frac{\beta}{2} R_c + \frac{\beta c_3}{2}\right)
\end{align}
we get the regularized action to be
\begin{align}
	I_n= \frac{n l^{2}_{dS}}{ 8G_{4}} \tilde{A}(2r_h^3+c_3)= \frac{n l^{2}_{dS}}{ 8G_{4}} \tilde{A} r_h (r_h^2+1)
\end{align}
with $\beta=1=-4\alpha$. Then the Renyi entropy gives
\begin{align}
	S_1= i \frac{l^{2}_{dS}}{ 4G_{4}} \tilde{A}
\end{align}
with
\begin{align}
	\tilde{A}= \int dH_2= \int d\psi \int_{\delta}^{\frac{\pi}{2}} \frac{d\theta \cos(\theta)}{\sin[2](\theta)}= 2 \pi (\frac{1}{\delta}-1) \approxeq \frac{2\pi R_c}{l_{dS}}
\end{align}
Thus we see that the imaginary part of eq.\eqref{impart2} agrees with the above result just as in $dSW_4/CFT_2$.

\section{Partition Function in $DS/dS$ band}
\label{apppartfun}
The bulk metric for this case is given by
\bea \label{metric-DSd}
& & \hskip -0.3in ds^2=dw^2+\sin^2 \left(\frac{w}{l_{dS}}\right) ds_{d-1}^2=dw^2+\sin^2 \left(\frac{w}{l_{dS}}\right) \left(d\alpha^2+{l_{dS}^2}\sin^2 \left(\frac{\alpha}{l_{dS}}\right) ds_{d-2}^2\right).
\eea
The induced metric on EOW branes, $Y_{1,2}$ [$w=w_{1,2}$ slice in (\ref{metric-DSd})] has the following form
\bea \label{EOW-branes-DSd}
& & ds_{Y_{1,2}}^2=\sin^2 \left(\frac{w_{1,2}}{l_{dS}}\right) \left(d\alpha^2+{l_{dS}^2}\sin^2 \left(\frac{\alpha}{l_{dS}}\right) ds_{d-2}^2\right).
\eea
The dual CFT in this case lives at $\Sigma$ located at $\alpha \rightarrow \alpha_* ={{\pi l_{dS}} \over {2}}$. Note that $\Sigma$ is the DS/dS central slice of the $(d-1)$-dimensional brane theory, not an intersection of $Y_1$ and $Y_2$. In the bulk, the central slices of all intermediate spheres form the hypersurface $\{X_d=0\}\cap\mathcal B\simeq S^{d-2}\times[w_1,w_2]$ (purple in Fig.~\ref{fig:band}a).

Other quantities relevant to the computation of the Euclidean on shell action are given by
\bea \label{R-etc-DSd}
& &  R=\frac{d(d-1)}{l_{dS}^2} , \ \, \Lambda=\frac{(d-1)(d-2)}{2 l_{dS}^2}, \nn \\
& & K^{(1,2)}= \frac{d-1}{l_{dS}}\cot\left(\frac{w_{1,2}}{l_{dS}}\right) , \ \,T^{(1,2)} = \frac{d-2}{l_{dS}}\cot\left(\frac{w_{1,2}}{l_{dS}}\right).
\eea
The Euclidean on shell action can be obtained by simplifying the following gravitational action 
\bea \label{action-WH-ii-DS4}
& &\hskip -0.4in I_{dS {\mathcal B}_d}=-\frac{(d-1)}{8\pi G_d\ldS^2}\int_{w_1}^{w_2}d^dx \sqrt g+\frac{\cot\left(\frac{w_1}{\ldS}\right)}{8\pi G_d\ldS}\int_{Y_1} d^{d-1}x \sqrt h-\frac{\cot\left(\frac{w_2}{\ldS}\right)}{8\pi G_d\ldS}\int_{Y_2} d^{d-1}x \sqrt h.
\eea
For four-dimensional de Sitter space as a bulk theory, i.e., $d=4$, metric (\ref{metric-DSd}) is simplified to the following form
\bea \label{metric-DS4}
& & \hskip -0.25in
ds^2=dw^2+\sin^2 \left(\frac{w}{l_{dS}}\right) \left(d\alpha^2+{l_{dS}^2}\sin^2 \left(\frac{\alpha}{l_{dS}}\right) ds_{2}^2\right)\nn \\
& & =dw^2+\sin^2 \left(\frac{w}{l_{dS}}\right) \left(d\alpha^2+{l_{dS}^2}\sin^2 \left(\frac{\alpha}{l_{dS}}\right) (d\theta^2+\sin^2\theta d\phi^2)\right).
\eea 
The band is described as $w_1 \leq w \leq w_2$ and the range of $\alpha$ is: $\alpha \in [0, \pi l_{dS}]$. We perform the $\alpha$ integral for the general range $\alpha_1$ to $\alpha_2$ and then fix the required values of $\alpha_{1,2}$. In this case,  on shell action (\ref{action-WH-ii-DS4}) for the metric (\ref{metric-DS4}) simplifies to
\bea \label{action-WH-iii-DS4}
 \hskip -0.25in I_{dS\mathcal B_4}
&&=-\frac{{\rm Vol}(S^2)\ldS}{8\pi G_4} 2\left(\cos\left(\frac{w_1}{\ldS}\right)-\cos\left(\frac{w_2}{\ldS}\right)\right) I_\alpha,
\eea
where
\bea \label{Iw-Ialpha}
& &  \hskip -0.3in  I_{\alpha}=\int_{\alpha_1}^{\alpha_2} d\alpha \sin^2 \left(\frac{\alpha}{l_{dS}}\right)=\frac{1}{4} \left(l_{dS} \left(\sin \left(\frac{2 \alpha _1}{l_{dS}}\right)-\sin \left(\frac{2
	\alpha _2}{l_{dS}}\right)\right)-2 \alpha _1+2 \alpha _2\right).
\eea
For $\alpha_1=0$ and $\alpha_2=\frac{\pi l_{dS}}{2}$, $I_{\alpha}=\frac{\pi l_{dS}}{4}$; for the full brane, {$\alpha_1=0$} and {$\alpha_2=\pi\ldS$}, {$I_\alpha=\frac{\pi\ldS}{2}$}; and for $\alpha_1=0$ and $\alpha_2=\alpha_*$, $I_{\alpha}=-\frac{l_{dS}}{4} \sin \left(\frac{2
	\alpha _*}{l_{dS}}\right)+\frac{\alpha _*}{2}$. Using ${\rm Vol}(S^2)=4 \pi$ and (\ref{eq:G3F}), the on-shell action (\ref{action-WH-iii-DS4}) is further simplified as
\begin{equation}\label{eq:IWE3}
	I_{dS{\mathcal B}_4}=-\frac{\ldS}{2G_4}\hat F_wI_\alpha=-\frac{I_\alpha}{G_3}.
\end{equation}
Since the $\alpha$ integral runs over a compact $S^3$, the action is finite: there is no UV divergence and hence no logarithmic term. 

Instead, for the full brane ($\alpha\in[0,\pi\ldS]$) we find
\begin{equation}
	I_{dS{\mathcal B}_4}=-\frac{\pi\ldS}{2G_3},\qquad Z_{CFT}\sim e^{-I_{dS{\mathcal B}_4}}=e^{\frac{\pi\ldS}{2G_3}}=e^{S_{dS_3}},
\end{equation}
i.e. $\log Z$ equals the de Sitter entropy of the $dS_3$ brane theory. In $DS_3/dS_2$ this entropy is related to the central charge of the dual pair of CFTs by $S_{dS_3}=\frac{\pi c}{3}$ \cite{Dong:2018cuv,Geng:2019ruz}, which gives
\begin{equation}\label{eq:cE}
	c=\frac{3\ldS}{2G_3}=\frac{3\ldS^2}{4G_4}\hat F_w.
\end{equation}
Hence, the central charge is real in this case and coincides with the central charge of two dimensional CFT in $DS_3/dS_2$ correspondence, using the same identification (\ref{eq:G3F}) of Newton constants as in all other sections.

	\noindent{\bf Reduction for general $d$.} The above analysis can be extended to any dimension. Write $V_{d-1}\equiv{\rm Vol}(S^{d-1}_{l_{dS}})$. Then the measures are $\sqrt g\,d^dx=l_{dS}\,d\hat w\,\sin^{d-1}\!\hat w\,dV_{d-1}$ and $\sqrt h=\sin^{d-1}\!\hat w_\beta\,dV_{d-1}$, and (\ref{action-WH-ii-DS4}) becomes
	\be
	I_{dS{\mathcal B}_d}=-\frac{V_{d-1}}{8\pi G_d\,l_{dS}}\Big\{(d-1)\int_{\hat w_1}^{\hat w_2}\sin^{d-1}\hat w\,d\hat w+\Big[\sin^{d-2}\hat w\cos\hat w\Big]_{\hat w_1}^{\hat w_2}\Big\}.
	\ee
	The identity
	\[
	\frac{d}{d\hat w}\big(\sin^{d-2}\hat w\cos\hat w\big)=(d-2)\sin^{d-3}\hat w-(d-1)\sin^{d-1}\hat w
	\]
	cancels the $\sin^{d-1}$ terms between bulk and branes, leaving
	\be\label{band-Gd}
	I_{dS{\mathcal B}_d}=-\frac{(d-2)\,V_{d-1}}{8\pi G_{d-1}\,l_{dS}^2},\qquad\frac{1}{G_{d-1}}=\frac{l_{dS}}{G_d}\int_{\hat w_1}^{\hat w_2}\sin^{d-3}\hat w\,d\hat w .
	\ee
	This is exactly the Euclidean on-shell action of $(d-1)$-dimensional de Sitter gravity on $S^{d-1}_{l_{dS}}$; for $d=4$ it reproduces (\ref{eq:IWE3}). The direction $w$ enters only through $G_{d-1}$.
		
	\noindent{\bf Brane-removal check: nested DS/dS.} Let $w_1\to0$ and $w_2\to\pi l_{dS}$. The brane terms vanish, since $\sin^{d-2}\hat w_\beta\cos\hat w_\beta\to0$, and the band becomes all of $S^d$ (Fig.~\ref{fig:band}b). Using ${\rm Vol}(S^d)={\rm Vol}(S^{d-1})\,l_{dS}\int_0^\pi\sin^{d-1}\hat w\,d\hat w$ and $\int_0^\pi\sin^{d-1}=\frac{d-2}{d-1}\int_0^\pi\sin^{d-3}$, (\ref{band-Gd}) gives
	\be\label{band-check}
	I_{dS{\mathcal B}_d}\ \to\ -\frac{(d-1)\,{\rm Vol}(S^d_{l_{dS}})}{8\pi G_d\,l_{dS}^2}=I_{S^d},\qquad\log Z=S_{dS_d}.
	\ee
	This is the Euclidean action of the full $S^d$, and $\log Z$ is the de Sitter entropy \cite{Gibbons:1977mu} in $d$ dimensions. For $d=4$: $1/G_3=2l_{dS}/G_4$, $\log Z=\pi l_{dS}/(2G_3)=\pi l_{dS}^2/G_4=S_{dS_4}$, and $c=3l_{dS}/(2G_3)=3l_{dS}^2/G_4$. Without branes the construction is therefore DS/dS applied twice, $DS_4\to dS_3\to dS_2$. With branes, it is $DS_d/dS_{d-1}$ in which both throats near $w\to0,\pi l_{dS}$ (associated with the ultraviolet of the two DS/dS CFT sectors~\cite{Alishahiha:2005dj}) have been replaced by Neumann branes. This is reminiscent of cut-off DS/dS and its $T\bar T$ description~\cite{Gorbenko:2018oov}, but with Neumann rather than Dirichlet conditions; the precise relation is left open.
	

 \section{$DS/dS$ band above the energy scale $\frac{1}{l_{dS}}$} 
\label{DS-dS-above-1/lds}
In $DS/dS$ duality, central charge of dual theory is zero if we are above the scale $\frac{1}{l_{dS}}$ \cite{Alishahiha:2005dj} due to the combined effect of $d-1$ dimensional gravity, matter and their interactions between the scale  $\frac{1}{l_{dS}}$ and the bulk Planck scale $M_d$. So far, what we discussed is that a pair of CFTs in $d-1$ dimensions are dual to de Sitter space in $d$ dimensions up to the energy scale $\frac{1}{l_{dS}}$. The gravity description is valid up to the Planck scale $M_d$. One natural question that one can ask is what the CFT interpretation is of gravity in the regime $\frac{1}{l_{dS}}<E < M_d$ or $M_d \gg \frac{1}{l_{dS}}$, i.e., analysis of the lower dimensional field theory and gravity description in the energy range $\frac{1}{l_{dS}}<E < M_d$ using bulk gravity. The effective operator dimensions and the effective central charge decrease dramatically for $E \gg \frac{1}{l_{dS}}$. These effects and the other related effects suggest gravitational dressing and cancellation of the conformal anomaly in the $d-1$ dimensional theory at energies above $\frac{1}{l_{dS}}$. Let us discuss this in detail.

$DS/dS$ band in terms of $z$ coordinate is described as follows. We take the bulk de Sitter space in $d$ dimensions and then localize this on $d-1$ dimensional EOW branes located at $z=z_{1,2}$ where $z_{1,2}=z(w_{1,2})$. Next, we dualize $d-1$ dimensional de Sitter gravity using $DS/dS$ correspondence, and the dual CFT lives at $\tilde{z}=0$ where $\tilde{z}$ is the coordinate of $dS_{d-1}$ gravity on EOW branes. The coordinates $z$ and $\tilde{z}$ will appear soon. Let us rewrite the metric for de Sitter space in $d$ dimensions in terms of the $z$ and $w$ coordinates as follows
\bea \label{dS-dS-r}
& & ds^2=\frac{1}{\cosh^2\left(\frac{z}{l_{dS}}\right)}\left(dz^2+ds^2_{dS_{d-1}}\right)= dw^2+\sin^2\left(\frac{w}{l_{dS}}\right)ds^2_{dS_{d-1}}.
\eea
Relationship between $z$ and $w$ using (\ref{dS-dS-r}) is obtained as
\bea \label{z(w)}
& & z(w)=l_{dS}\cosh^{-1}\left(\frac{1}{\sin\left(\frac{w}{l_{dS}}\right)}\right).
\eea
The central slice $w=\frac{\pi l_{dS}}{2}$ corresponds to $z=0$. Let us focus on the bulk term in the on-shell action (\ref{action-WH-ii-DS4})
\bea \label{OS-z}
& & I_{dS{\mathcal B}_4}=-\frac{3 }{8 \pi G_4 l_{dS}^2}\int_{z_1}^{z_2}dz\, d^{d-1}x\sqrt{g}
\eea
The metric for the de Sitter space in four dimensions in terms of $z$ coordinate is written below
\bea \label{metric-DS4-z}
& & \hskip -0.4in
ds^2=\frac{1}{\cosh^2\left(\frac{z}{l_{dS}}\right)}\left[dz^2+ds^2_{dS_{3}}\right] =\frac{1}{\cosh^2\left(\frac{z}{l_{dS}}\right)}\left[dz^2+\frac{1}{\cosh^2\left(\frac{\tilde{z}}{l_{dS}}\right)} \left(d\tilde{z}^2+ds_{dS_2}^2\right)\right].
\eea 
Comparing the second line of (\ref{metric-DS4-z}) with (\ref{metric-DS4}), we find that metric for $dS_3$ slices in terms of $\tilde{z}$ and $\alpha$ coordinates are obtained as follows
\bea \label{dS3-ztilde-alpha}
& & ds^2_{dS_3}=\frac{1}{\cosh^2\left(\frac{\tilde{z}}{l_{dS}}\right)} \left(d\tilde{z}^2+ds_{dS_2}^2\right) =d\alpha^2+{l_{dS}^2}\sin^2 \left(\frac{\alpha}{l_{dS}}\right) {d\hat s^2_2}\,
\eea
where $ds^2_{dS_2}=\ldS^2d\hat s^2_2$ and $d\hat s^2_2$ is the unit radius metric. Using (\ref{dS3-ztilde-alpha}), $\tilde{z}$ in terms of $\alpha$ is given by
\bea
& &\tilde{z}(\alpha)=l_{dS}\cosh^{-1}\left(\frac{1}{\sin \left(\frac{\alpha}{l_{dS}}\right)}\right).
\eea
Again the central slice, $\alpha=\alpha_*=\frac{\pi l_{dS}}{2}$ corresponds to $\tilde{z}=0$. Including the brane terms, the wedge action reduces to the three-dimensional action, see (\ref{actdswh}). On shell action for four-dimensional de Sitter space (\ref{metric-DS4-z}) as a bulk theory is obtained as follows
\begin{align} \label{OS-DS4-z}
I_{dS{\mathcal B}_4}&\sim-\frac{V_2}{4\pi G_3\ldS^2}\int_{\tilde z_1}^{\tilde z_2}\frac{d\tilde z}{\cosh^3\left(\frac{\tilde z}{\ldS}\right)}\nonumber\\
&=-\frac{V_2}{4\pi G_3\ldS^2}\left[\frac12\ldS\tan^{-1}\left(\sinh\left(\frac{\tilde z}{\ldS}\right)\right)+\frac12\ldS\tanh\left(\frac{\tilde z}{\ldS}\right){\rm sech}\left(\frac{\tilde z}{\ldS}\right)\right]_{\tilde z_1}^{\tilde z_2},
\end{align}
where $V_2$ is the volume of the two-dimensional slice and the effective Newton constant obtained by dimensional reduction is
\begin{equation}
\frac{1}{G_3}=\frac{1}{G_4}\int_{z_1}^{z_2}\frac{dz}{\cosh^2\left(\frac{z}{\ldS}\right)}=\frac{\ldS}{G_4}\left(\cos\hat w_1-\cos\hat w_2\right),
\end{equation}
in agreement with main text (using $dw=dz/\cosh(z/\ldS)$). 
Near $\tilde{z}=0$, we have
\bea
& & \int \frac{d\tilde{z}}{\cosh^3\left(\frac{\tilde{z}}{l_{dS}}\right)} \sim \tilde{z}-\frac{\tilde{z}^3}{2 l_{dS}^2}+O(\tilde{z}^5).
\eea

Hence, there is no log term in (\ref{OS-DS4-z}). This is consistent with the vanishing of the conformal anomaly in the energy regime $\frac{1}{l_{dS}}<E < M_d$ found in \cite{Alishahiha:2005dj}. We stress, however, that the absence of a logarithm is by itself only suggestive. Indeed, with $\sin(\alpha/l_{dS})={\rm sech}(\tilde z/l_{dS})$ the integral in (\ref{OS-DS4-z}) is the same as $I_\alpha=\int d\alpha\,\sin^2(\alpha/l_{dS})$ of section \ref{apppartfun}, from which the finite action yields $c=\frac{3l_{dS}}{2G_3}\neq 0$ through $S_{dS_3}=\frac{\pi c}{3}$. The absence of a logarithm therefore shows only that the central charge cannot be read off from an anomaly-type divergence. Demonstrating that the effective central charge vanishes for $\frac{1}{l_{dS}}<E<M_d$ would require the analogue of the varying-Newton-constant construction of \cite{Alishahiha:2005dj}  in $DS/dS$ band.


		\vspace{-0.6cm}

\newpage
\bibliographystyle{JHEP}
\bibliography{refs}

	\end{document}